\documentclass[onecolumn,11pt]{article}
\usepackage[top=1in, bottom=1in, left=1in, right=1in]{geometry}
\usepackage{amsmath,amsfonts,amscd,amssymb}
\usepackage{subcaption}
\usepackage[section]{placeins} 

\usepackage{url}
\usepackage{color}
\usepackage{rotating}
\usepackage{multirow}
\usepackage{wrapfig}
\usepackage{mathtools}
\usepackage{subeqnarray}
\usepackage{setspace}
\usepackage[numbers]{natbib}
\usepackage{graphicx}

\DeclareMathOperator{\T}{\mathbf{T}}
\DeclareMathOperator{\X}{\mathbf{X}}
\DeclareMathOperator{\W}{\mathbf{W}}
\DeclareMathOperator{\C}{\mathbf{C}}
\DeclareMathOperator{\B}{\mathbf{B}}
\DeclareMathOperator{\U}{\mathbf{U}}

\title{Data-driven Modeling of Two-Dimensional Detonation Wave Fronts}

\author{Ariana Mendible$^1$\footnote{Corresponding author: mendible@uw.edu, 3900 E Stevens Way NE, Seattle, WA 98195} , Weston Lowrie$^2$,\\ Steven L. Brunton$^1$, and J. Nathan Kutz$^3$\\
{\small \em $^1$Department of Mechanical Engineering, University of Washington, Seattle, WA 98195} \\
{\small \em $^2$Applied Research Associates, Inc. Seattle, WA} \\
{\small \em $^3$Department of Applied Mathematics, University of Washington, Seattle, WA 98195}}

\date{\today}

\begin{document}

\maketitle

\begin{abstract}
    Historical experimental testing of {\em high-altitude nuclear explosions} (HANEs) are known to cause severe and detrimental effects to radio frequency signals and communications infrastructure. In order to study and predict the impact of HANEs, tractable computational approaches are required to model the complex physical processes involved in the detonation wave physics. Modern reduced-order models (ROMs) can enable long-time and many-parameter simulations with minimal computational cost. However, translational and scale invariances inherent to this type of wave propagation problem are known to limit traditional ROM approaches. Specifically, dimensionality reduction methods are typically ineffective in producing low-rank models when invariances are present in the data.  In this work, an unsupervised machine learning method is used to discover coordinate systems that make such invariances amenable to traditional dimensionality reduction methods. The method, which has previously been demonstrated on one-dimensional translations, is extended to higher dimensions and additional invariances. A surrogate HANE system, i.e. a HANE-ROM, with one detonation wave is captured well at extremely low-rank. Two detonation-waves are also considered with various amounts of interaction between the waves, with improvements to low-rank models for multiple wave quantities with limited interaction. 
\end{abstract}

\section{Introduction}

Reduced-order models (ROMs) aim to alleviate the computational burden of simulating high-dimensional systems that often arise from the discretization of partial differential equations. 
ROMs rely on modal decompositions to compute a set of low-rank basis modes~\cite{Taira2017aiaa,taira2020modal}, a small number of which can be used to approximate the full spatiotemporal dynamics~\cite{benner2015survey,Kutz:2013,Brunton2019book}. Such modal decompositions have been extensively studied and have proven to be efficient computational proxies for the original governing equations~\cite{benner2015survey,antoulas2005approximation,Taira2017aiaa,taira2020modal,quarteroni2015reduced,hesthaven2016certified}.  The diversity of modal decompositions include the {\em dynamic mode decomposition} (DMD)~\cite{Schmid2010jfm,Rowley2009jfm,Tu2014jcd,Kutz2016book}, {\em principal components analysis} (PCA) and its variants~\cite{Candes:2011}, and {\em proper orthogonal decomposition} (POD) and its variants~\cite{Towne2018jfm, willcox2006cf,rowley2005model,berkooz1993proper}. The \textit{singular value decomposition} (SVD) is central to all of these dimensionality-reduction approaches, computing the optimal low-rank basis that is tailored to the specific data set. The SVD, however,  assumes the separation of space and time variables. In many, if not most cases, this assumption does not hold because the dynamics are inherently coupled in space and time, including the well-known invariances of translation, rotation, and scaling.  This impairs the effectiveness of traditional dimensionality reduction methods because of this inseparability of space and time~\cite{Brunton2019book,kirby1992reconstructing,Rowley2000physd,Rim2018juq,Reiss2018jsc}.  In this manuscript, we propose a data-driven algorithm for the discovery of coordinate systems in which we can explicitly track and remove invariances from spatio-temporal data, thus allowing for improved ROM construction.

\begin{figure}[t!]
	\centering
	\begin{subfigure}[b]{0.35\linewidth}
		\includegraphics[width = \linewidth]{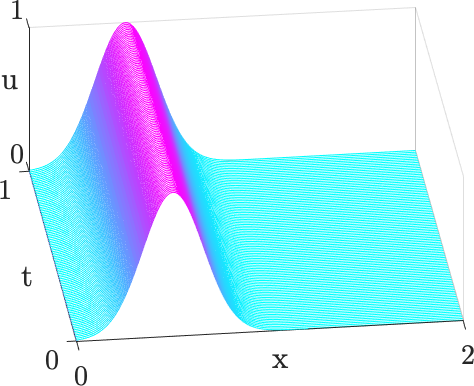}
		\caption{stationary}
	\end{subfigure}
	\qquad
	\begin{subfigure}[b]{0.35\linewidth}
		\includegraphics[width = \linewidth]{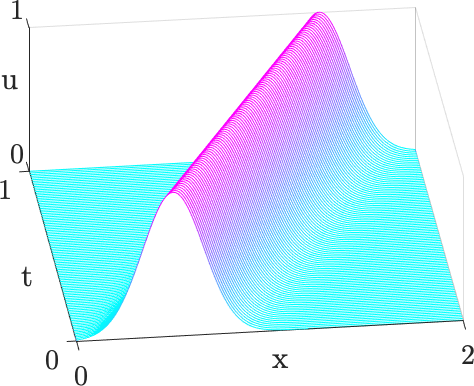}
		\caption{traveling}
	\end{subfigure}
	\begin{subfigure}[b]{0.35\linewidth}
		\includegraphics[width = \linewidth]{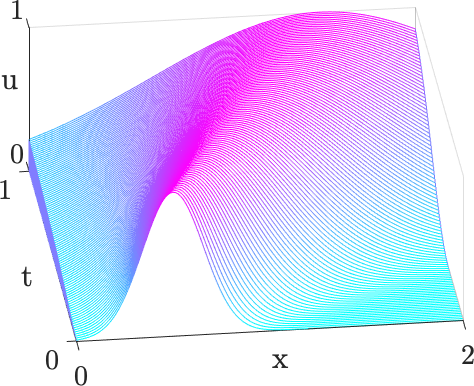}
		\caption{traveling \& growing}
	\end{subfigure}
	\qquad
	\begin{subfigure}[b]{0.35\linewidth}
		\includegraphics[width = \linewidth]{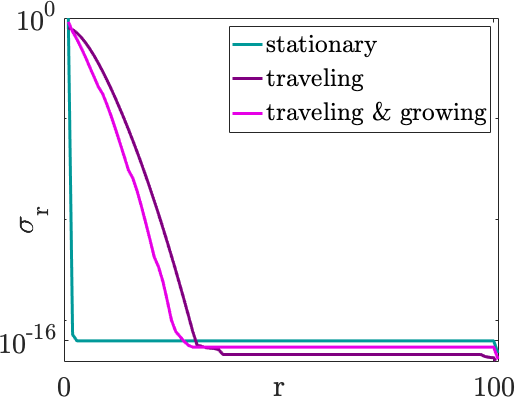}
		\caption{Singular Values}
	\end{subfigure}
	\caption{Three levels of invariance imposed to a soliton wave. A stationary wave contains no time-symmetries, the traveling wave exhibits a translational invariance, and the traveling \& growing wave contains a translational and scaling invariance. Singular value spectra for each clearly show that these invariances inhibit the effectiveness of the dimensionality reduction-- many more modes are needed to approximate the symmetrical data.}
	\label{fig:fig1}
\end{figure}

To illustrate how invariances undermine traditional ROMs, consider the three traveling wave systems shown in Figure \ref{fig:fig1}.  Figure \ref{fig:fig1}a shows a stationary, localized wave, with a constant wave shape in time and space. Figure \ref{fig:fig1}b shows the simplest wave motion possible: a wave traveling in time with a constant wave speed. Figure \ref{fig:fig1}c shows the same traveling wave also growing, or scaling, with time. The singular value decay of each data set is given in Figure \ref{fig:fig1}d. Ideally, the singular values would decay rapidly for a good low-rank embedding, with only a few modes needed to capture the dynamics. While the SVD of the stationary wave captures this well, with only one mode containing the vast majority of the data variance or pulse energy, the SVDs of the data with invariance (translation and scaling) show the opposite. Here, the energy is spread into many modes, meaning more modes must be retained in order to approximate the system well. In this case dimensionality reduction is inefficient and downstream computational tasks will remain expensive. Specifically, the dynamics for each case should be of rank-1, -2 (with translation) and -3 (with translation and scaling) respectively.  Instead, the nonstationary data with invariances requires nearly a rank-40 approximation, which is more than an order-of-magnitude higher in dimension.  This artificial increase in dimension is exactly what we aim to circumvent with our algorithms.

In this work, we seek to ameliorate the challenges posed to dimensionality reduction by such invariances. First, the existing methods for dimensionality reduction under translational invariances, or wave motion, will be discussed. The dimensionality reduction method developed in previous work will be introduced, and extensions will be presented to include higher dimensions and a wider variety of symmetries. The success of the augmented method will be verified by a simple example. Then, a two-dimensional multiphysics example of HANE will be explored, using an idealized system to study the effectiveness of the dimensionality reduction. Comparisons of the method to traditional methods are presented, and further improvements to the method will be discussed. This work aims to accomplish three primary goals (1) develop a method to address translational and scaling invariances in dimensionality reductions for higher-dimensional data, (2) discover interpretable and physically relevant models for these invariances and the wave shapes, and (3) explore effectiveness of the method on dimensionality reduction of complex physical systems.

\section{Background}
\subsection{Wave Motion in Reduced-Order Models}
The issues posed by translational invariance within spatiotemporal data sets are well known to hinder the effectiveness of SVD-based models. While these methods are still widely-used, there is growing interest in developing mathematical architectures that are capable of determining the traveling wave frame of reference of the underling wave~\cite{kirby1992reconstructing,Rowley2000physd,carlberg2015adaptive,Rim2018juq,Reiss2018jsc}. Indeed, a common focus of current literature is finding a translational coordinate frame in which traveling waves appear stationary. These shift-based approaches utilize a range of techniques to discover such coordinate frames, many relying on an approximately-constant wave shape or sharp, identifiable shock front for speed determination. While these methods are critical to address the shortcomings of traditional methods, they are limited to applications with constant wave speeds or knowledge of the underlying physics, they lack interpretability, and they are often limited to one spatial dimension or one-dimensional shifts. In addition, many of these methods do not consider additional invariances, such as scaling and/or rotation. Recently, an unsupervised machine learning procedure has been proposed for transport-dominated systems characterized by traveling waves~\cite{mendible2020dimensionality,mendible2021}. This data-driven method, termed the {\em Unsupervised Traveling Wave Identification with Shifting and Truncation} (UnTWIST), can be applied with or without knowledge of the governing equations, providing an interpretable mathematical framework for ROMs exhibiting traveling wave phenomenon.  The UnTWIST method has been developed for translational symmetries and has been demonstrated on one-dimensional phenomenon. Here, UnTWIST is generalized to include additional symmetries and incorporate higher spatial (e.g. 2D and 3D) dimensional data. 

\subsection{High-Altitude Nuclear Explosions}

HANEs are nuclear explosions that detonate within Earth's atmosphere above an altitude of approximately 30 kilometers. Historically, HANE testing was conducted between 1958 and 1962 to determine the collateral effects of such explosions. There is limited data and information known about the wide-reaching multi-physics effects of HANEs from these historical experimental observations~\cite{hess1964effects}. Ionization environments in Earth's upper atmosphere following HANEs impact a variety of military and civilian systems, including radio frequency (RF) signal propagation used for communications, radar, and GPS~\cite{hoerlin1976united, arendt1964effects}. However, because of the obvious drawbacks of experimental testing and the immense threat posed by such weapons, it is critical to model and predict the behavior and effects of HANEs via computation.  To accurately predict impacts of HANEs on these systems, typically high-fidelity, multi-physics computational models are employed. These large-scale and computationally expensive simulation models include hydrodynamic systems of equations, such as the Euler system.  Investigating a diversity of HANE scenarios is typically computationally intractable, thus motivating the development of ROMs.  A challenge for data-driven analysis of numerical models of hydrodynamic systems is that they often involve traveling waves, moving shocks, or sharp gradients and discontinuities. These types of physical features are exceptionally difficult to model in a principled manner.  However, a recent set of data-driven  algorithmic improvements have made them tractable.  Applying these machine learning innovations to HANEs is the focal point of this manuscript.

\section{Methods}

In what follows, the methods and algorithms developed for characterizing systems with invariances is outlined.  Beginning with establishing a simulation engine for the generation of appropriate spatiotemporal data, the subsequent sections develop the general mathematical innovations for extracting invariances from this data.

\subsection{Euler Equation Simulations}
While detailed HANE simulations are complex and include the modeling of multiple physical effects, an idealized test problem can capture some of the dominant aspects of high-altitude spatio-temporal dynamics.  For example, many aspects can be resolved by solving the Euler equations on a domain with a spherical perturbation of high temperature in a vertically stratified background atmosphere.  This idealized problem can provide enough complexity to test ROM algorithms while capturing important phenomenology such as the generation of radially expanding waves and gravitational buoyancy effects, resulting in translation and scaling invariances that must be handled concurrently.

The three-dimensional Euler equations are given by 
\begin{equation}\label{eq:3deuler}
	\frac{\partial}{\partial t} 
	\begin{bmatrix}
	\rho\\ \rho \mathbf{v} \\ E
	\end{bmatrix}
	+ \nabla \cdot
	\begin{bmatrix}
	\rho \mathbf{v} \\ \rho \mathbf{vv} + p\mathbf{I}\\ (E+p)\mathbf{v} 
	\end{bmatrix}
	 = 
	 \begin{bmatrix}
	 0\\ -\rho \mathbf{g}\\ 0
	 \end{bmatrix}
\end{equation}
where $\rho$ is the mass density, $\mathbf{v}$ is the velocity vector, $p$ is the pressure, $\mathbf{I}$ is the identity matrix, $E$ is the total energy defined by $E = {p}/({\gamma-1})+ p |\mathbf{v}|^2/2$, $\gamma$ is the ratio of specific heats, and $\mathbf{g}$ is the vector of acceleration due to gravity. 

The Euler system (\ref{eq:3deuler}) was simulated for time 0 to 160 seconds using the open-source PyClaw software package \cite{clawpack, mandli2016clawpack, pyclaw-sisc}, using the Euler with gravity Riemann solver.  A $\textrm{CFL}= 0.5$ is used with data outputs every 1 second.  The simulations use a spherical coordinate system domain 
\begin{align*}
	\theta &\in [\pi/4-0.15, \pi/4+0.15] = [\theta_-,\theta_+] \text{ rad} \\
	\phi &\in [0, 0.30] =[\phi_-,\phi_+] \text{ rad}  \\
	r &\in [80e5, 1559e5] = [r_-,r_+] \text{ cm} 
\end{align*}
which corresponds to about a 1000 km by 1000 km latitude and longitude extent and 80 to 1559 km altitude range.  A background mass density and pressure are initialized from tables to yield a vertically stratified background atmosphere.  The pressure is slightly modified to maintain hydrostatic equilibrium using the f-wave  method~\cite{BaleLevMitRoss02}.

Extrapolation boundary conditions are used in the $\theta_\pm$, $\phi_\pm$ and the top ($r_+$) boundaries.  At the bottom ($r_-$) boundary the initial background atmosphere conditions are held fixed
\begin{align*}
    \rho_{k=0} &= \rho_0 \\
    \rho \mathbf{v}_{k=0} &= 0 \\
    E_{k=0} &= \frac{p_0}{\gamma - 1} 
\end{align*}
where $\rho_0$ and $p_0$ are the initial background mass density and pressure and $k$ is the index for the radial $r$ dimension.

\begin{figure}[t!]
    \centering
    \begin{subfigure}{\linewidth}
	\includegraphics[width=\linewidth]{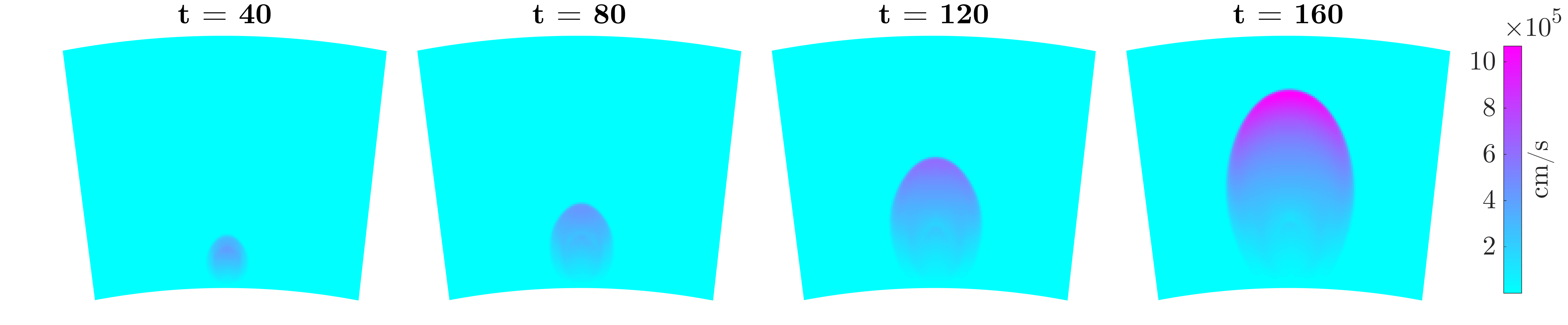}
    \end{subfigure}
    \hfill
    \begin{subfigure}{\linewidth}
	\includegraphics[width=\linewidth]{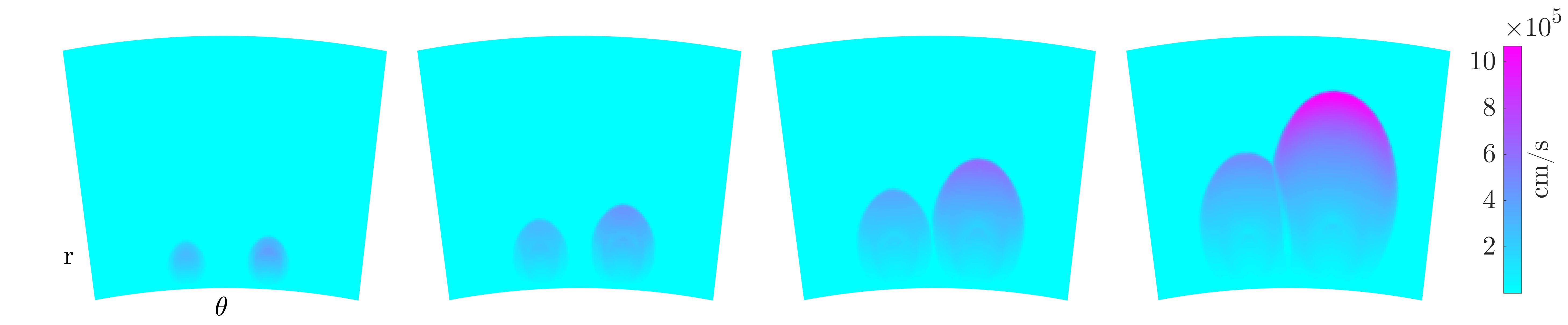}
    \end{subfigure}
    \caption[Example of one and two initial perturbation data sets]{Velocity magnitude of each type of data set: (top) one initial energy perturbation of varying magnitude, (bottom) two initial energy perturbations of a given magnitude at varying distances, each at four time steps as viewed from the central $\phi$ plane of the three-dimensional view.}
    \label{fig:example_data}
\end{figure}

Two main data sets are simulated using these equations. In the first set of simulations, a single spherical temperature perturbation is constructed with various initial energies, $E=$1.0, 1.6, and 2.0 eV, and propagated in time. In the second set of simulations two spherical temperature perturbations of 1.0 eV and 1.6 eV initial energy are simulated with various distances separating the two. The perturbations are initialized at an altitude of 200 km with a peak temperature at the center and falling off linearly until a radius of 100 km.  Figure~\ref{fig:example_data} shows the evolution dynamics for the two simulation scenarios considered here.  The underlying wave motion physics includes the propagation of a rising and expanding fluid field which we wish to characterize in a principled way using a ROM to discover the observed translational and scaling invariances.

\subsection{Unsupervised Traveling Wave Identification with Shifting and Truncation}
In the Euler system (\ref{eq:3deuler}) as a HANE approximation, there are three persistent invariances that may impede traditional dimensionality reduction approaches. The coherent structure is (1) traveling in the $r$-direction, (2) potentially traveling in the $\theta$-direction, and (3) growing in size. By viewing the data in a coordinate frame that makes the wave structure stationary and scale-invariant, low-dimensional approximations can capture the details within the waves, agnostic of position or size, rather than using many modes to compensate for this translation and scaling phenomenon. 

To learn models for these underlying invariances, the UnTWIST method will be generalized~\cite{mendible2020dimensionality}. UnTWIST is a recently-developed method that shows promising improvements in dimensionality reduction for complex wave shapes in both computational and experimental data~\cite{mendible2021}. This method, similar to other shift-based methods, learns a moving coordinate frame, given by the speed of a traveling wave. Transforming the data into this coordinate frame holds the wave of interest stationary, allowing for models to be built in the wave reference frame. Unique to UnTWIST is the ability to learn physically-relevant, interpretable wave speeds directly from multiple-wave data with little knowledge of underlying dynamics. Previously, the UnTWIST method has been formulated and demonstrated only for one-dimensional data. In this work, the method will be augmented to account for additional dimensions and symmetries. This generalization will not only allow its application on HANE data with the aforementioned two-dimensional translation and scaling, it will allow future exploration on higher-dimensional systems with other invariances. 

The method innovations will be introduced in two sections. The first describes the detection of the invariances inherent to the wave data. The second describes the setup of the optimization problem and how the resulting models are used to align and approximate data in a low rank manner.  Both sections highlight the critical extensions to UnTWIST which allow the optimization to generalize to a broader class of problems and in higher spatial dimensions.

\subsubsection{Invariance Detection}\label{subsec:initialization}

\begin{figure}[t!]
	\centering
		\includegraphics[width = \linewidth]{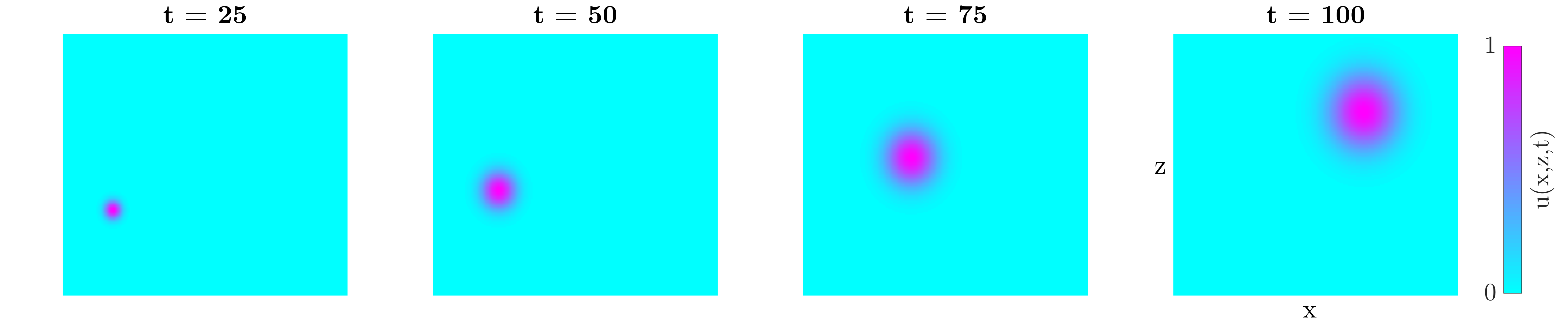}
	\caption[Time series of example problem]{Time series showing the wave height of an example Gaussian traveling in two dimensions and increasing in size, exemplifying the translational and scaling symmetries.}
	\label{fig:madeup}
\end{figure}

The first step in processing the data using UnTWIST is detecting the location of the wave structure, as well as the invariances we seek to account for. In general, there are many methods that could be utilized to detect these parameters, with computer vision and machine learning providing a suite of potential detection techniques. Here, a simple threshold-based detection of locations of each wave in the $x$- and $z$-dimensions and a scaling parameter is determined. These approaches can be generalized to a third spatial dimension, or additional rotational or other invariance if present. An example of the problems that will be considered here is given in Figure \ref{fig:madeup}.

\begin{figure}[t!]
	\centering
	\includegraphics[width=0.6\linewidth]{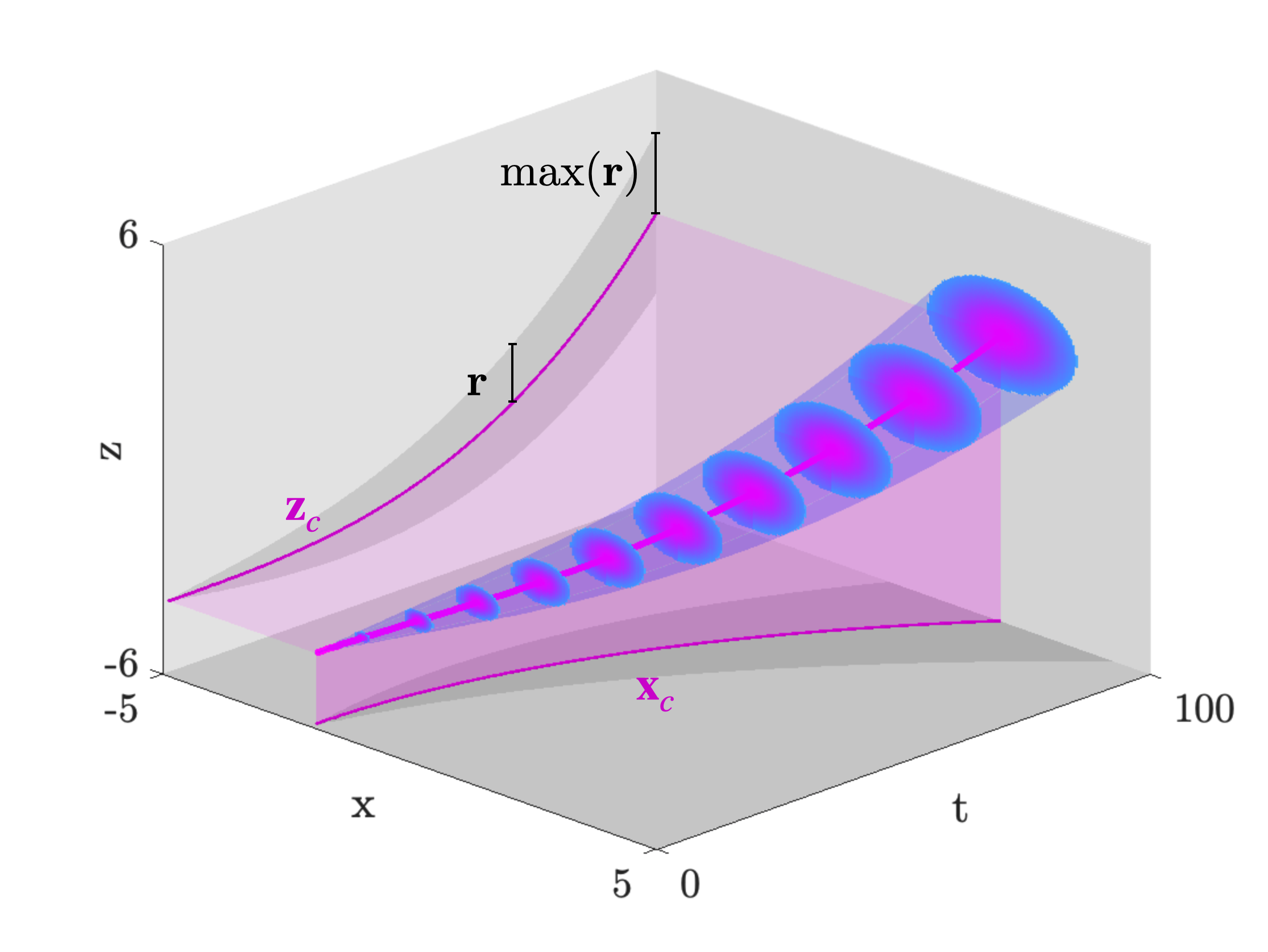}
	\caption[Schematic of invariance detection in higher dimensions]{Schematic of the shifting and scaling captured in the initialization step, with the same data shown in Figure \ref{fig:madeup}. Slices of the wave above the threshold, $\tau$, are plotted in time (disks). The center locations of the wave are shown piercing through the data slices in pink. The center locations are projected in pink to show the $\mathbf{x}_c$ and $\mathbf{z}_c$ in time. Radius of the wave in time $\mathbf{r}$ is also shown projected in gray, with the maximum radius in this example at the end of the time series.}
	\label{fig:madeup_cone}
\end{figure}

The input data is given as $\U \in \mathbb{R}^{m \times n \times q}$, where $\mathbf{x} \in i = 1:m$, $\mathbf{z} \in j = 1:n$, and $\mathbf{t}\in \ell = 1:q$. The first step in the wave detection is setting a simple threshold. In the examples that follow, a threshold $\tau$ is determined to be a certain percentile of wave height over the full data set 
\begin{equation}
	\U_\tau = \U> \tau
\end{equation}
For example, the $98^\text{th}$ percentile of values in $\U$ yields $\tau$, and all points above this $\tau$ will be considered within a quantity of interest. Time slices of those points within the $\U_\tau$ map can be seen as disks in Figure \ref{fig:madeup_cone}. 

Indices of the points that meet the threshold are then used to find the area of the quantity of interest as well as its $(x,z,t)$ location. 
The centers of the waves in the two dimensions, $\mathbf{x}_c$ and $\mathbf{z}_c$, shown as the pink lines projected on to the respective planes in Figure~\ref{fig:madeup_cone}, are given in cm by
\begin{equation}\label{eq:centers}
\begin{split}
\mathbf{x}_c(t_\ell) & = \frac{\sum_{i,j} \U_\tau (x_i,z_j,t_\ell) \mathbf{x}}{\sum_{i,j}\U_\tau (x_i,z_j,t_\ell)}\\
\mathbf{z}_c(t_\ell) & = \frac{\sum_{i,j} \U_\tau (x_i,z_j,t_\ell) \mathbf{z}}{\sum_{i,j}\U_\tau (x_i,z_j,t_\ell)}
\end{split}
\end{equation}
The shifts for these quantities are then computed as the distance from the center of the wave of interest $\mathbf{x}_c, \mathbf{z}_c$ to the center of the data set. For an input data set of size $\U(t_\ell) \in \mathbb{R}^{m \times n}$, the following two quantities, in pixels, are computed
\begin{equation}
\begin{split}
\text{shift}_x(t_\ell)  &= \lfloor (i(t_\ell) -m/2) \rceil \\
\text{shift}_z(t_\ell)  &= \lfloor (j(t_\ell) -n/2) \rceil
\end{split}
\end{equation}
where $\lfloor \cdot \rceil$ indicates rounding to the nearest integer, and ($i, j$) represent the indices within ($\mathbf{x},\mathbf{z}$) that correspond to centers ($\mathbf{x}_c, \mathbf{z}_c$). 
With these shift values, all waves will be centered within the frame at each time step, and will expand radially from this center point. 

The overall growth of the wave quantities must be counteracted in order to view the waves in stationary coordinate frames. To achieve this, an overall scaling parameter is determined based on the radius of the area of interest as approximated by a circle.
The area in cm$^2$, shown by the cone encompassing the disks in Figure \ref{fig:madeup_cone}, is computed by 
\begin{equation}
	\mathbf{a}(t_\ell) = dx  \cdot dz \cdot \sum_{i,j} \U_\tau(x_i,z_j,t_\ell). 
\end{equation}
The radius of each area of interest, shown by the gray area the $x-t$ and $z-t$ planes in Figure \ref{fig:madeup_cone} is computed as 
\begin{equation}\label{eq:radius}
\mathbf{r}(t_\ell) = \sqrt{\mathbf{a}(t_\ell)/\pi}.
\end{equation}
To retain as much detail in the data as possible, the scaling parameter is determined such that the largest original radius of the quantity of interest is scaled to unity, and all smaller radii are scaled up proportionally. The scaling parameter for each time step is then computed as
\begin{equation}
\mathbf{s}(t_\ell) = \frac{\max(\mathbf{r})}{\mathbf{r}(t_\ell)}.
\end{equation}
Three quantities, the shifts $\text{shift}_x(t_\ell) $ and $\text{shift}_z(t_\ell) $, as well as the scaling parameter $\mathbf{s}(t_\ell)$ are used as inputs to UnTWIST. The algorithm is modified to find sparse, interpretable models for these quantities. 

\subsubsection{Optimization}

The outcome of the invariance detection is a set of parameters, shift$_x$, shift$_z$ and $\mathbf{s}$, which can be used to align the wave shapes to be consistent throughout data. The optimization step now finds models for these parameters so that interpretable physics can be learned and accurate shifts and scaling can be applied. 
First, the individual invariance parameters must be separated by wave group. Given that many waves may be present in a data set, it is critical to create correlations, and resulting models, between parameters only for a given wave. These separations are initialized using a spectral clustering approach~\cite{ng2002spectral}, which robustly correlates continuous clusters. These grouping values, initialized in binary, can be used to mask off invariances as they apply to each wave in the data set.  
Once the wave fronts are identified and separated, the data is assembled into the optimization. 
A user-defined library of potential invariance functions is provided. Any number of library functions can be included, linear or nonlinear. Selecting functions which are likely to describe the particular wave invariances under consideration is practical, though a more complete basis can be provided at higher computational expense. 
The matrices $\X$ and $\T$ are constructed using the $(x_i,z_i,s_i,t_i)$ locations of the wave fronts within the data, where $\T$ contains the values of $t_i$ evaluated for each function in the user-defined library, and $\X$ contains the invariance values $(x_i,z_i,s_i)$. The cost function is given by Equation~\ref{eq:untwist}
\begin{equation}\label{eq:untwist}
\min_{\C, \B, \W \in \Omega}  \sum_{j = 1}^d \frac{1}{2} \W \odot \| \X_j-\T \C_j \|^2_2 + \lambda R(\B_j) + \frac{1}{2\zeta} \| \C_j-\B_j \|^2_2, 
\end{equation}
where $\W$ is the weighting matrix that serves to mask wave peak points for clustering into wave groups. With values of $0$ or $1$, each row of $\W$ corresponds to each wave invariance point $(x_i,z_i,s_i,t_i)$, and each column corresponds to a given wave. Values in $\C$ are the coefficients of the models that are discovered for each wave. A row of $\C$ corresponds to a wave, and columns give the coefficients of each term in the model library $\T$, which multiply together to generate the wave speed models. The sum in $j$ corresponds to the number of invariances with models to be fit, for example, for $(x_i,z_i,s_i)$, $d=3$. It is desirable for $\C$ to be sparse, i.e. to have few nonzero terms, to glean an interpretable, physically realistic model for the wave speeds. Rather than placing a sparsity constraint on $\C$ directly, the constraint can be relaxed by introducing an auxiliary matrix $\B$, which is close to $\C$. $\B$ is directly forced to be sparse via a regularizing function $R(\cdot)$, relieving the burden on $\C$ to meet both sparsity and accuracy goals. The hyperparameter $\lambda$ is chosen to calibrate the sparseness of an auxiliary matrix. The hyperparameter $\zeta$ is chosen to enforce the closeness of $\C$ and $\B$, ensuring that the solution $\C$ itself is also sparse. These two hyperparameters are tuned in tandem in order to meet sparsity and accuracy requirements of the model. This optimization presents a large search space over multiple parameters, and is not guaranteed to be convex. Sparse relaxed regularized regression~\cite{zheng2018unified} is used to minimize the cost function because of its ability to handle non-convexity and its computational efficiency compared to similar sparsity-promoting optimization schemes.

\subsection{Aligning the Invariances}
Once the model coefficients $\C$ are learned, they can be used to shift and scale each time segment of data in order to align the data into one wave group's moving coordinate frame. Smooth, equation-based invariance values can be reconstructed from the values in $\C$ and the model library. The shifts can be accounted for using a periodic shift operator, here using \texttt{circshift} in \textsc{Matlab}, separately in $x$ and $z$. The wave quantity of interest is shifted to the center of the data set. The scaling parameter is then used to rescale the data, with the largest-area quantity in the original data kept at its original scale. All other time slices are upsampled using \texttt{imresize} in \textsc{Matlab}. Other upsampling or resizing approaches can be used; the field of super-resolution is abundant with approaches for this step, but is outside the scope of this work.
These shifting and scaling procedures are performed for each wave quantity in the data separately, yielding $n$ frames which are then ready for traditional ROM modal decompositions. Indeed, once the data is aligned into the new coordinate frame, traditional dimensionality reduction methods can be applied and can be expected to reveal extremely low-rank modes for the aligned wave or wave group.


\section{Results}

\subsection{Simplified Example System}
To verify that the generalized version of the UnTWIST algorithm can indeed discover true, interpretable, low-rank underlying physics, an example problem with prescribed shifting and scaling is used as a benchmark. 

\paragraph{Example Data.}

In this example, a two-dimensional Gaussian ansatz is constructed as
\begin{equation}\label{eq:2dgaussian}
u(x,z,t) = \exp{\left( \frac{ -(x-c_x(t))^2-(z-c_z(t))^2 }{2t^2}\right) }
\end{equation}
with speeds in the $x$ and $z$ directions given by
\begin{equation}\label{eq:2dspeeds}
\begin{split}
c_x(t) &= 4t^2-2
\\
c_z(t) &= 6t^3-4
\end{split}
\end{equation}
respectively, in the domain $x = [-5:0.1:5]$, $z = [-6:0.1:6]$, and $t = [0.1:0.1:1]$. This example is shown in Figure \ref{fig:madeup}. These prescribed speeds, written in sparse, identifiable terms will be used to compare to the models resulting from UnTWIST. 

In order to find the centers of the Gaussian, a threshold $\tau$ is assigned to be the $98^{\text{th}}$ percentile of the total data. Here, $\tau = 0.32$. Solving for the radius of the intersection between the Gaussian and this threshold, we find
\begin{equation}\label{eq:trueradius}
	r =t \sqrt{-2\ln(\tau)}\approx 1.5 t. 
\end{equation}
This is the true physical radius that bounds the data we wish to consider. It will be used to compare to the outcome of the UnTWIST model. 

\paragraph{UnTWIST Results.} 

\begin{figure}[t!]
	\centering
	\includegraphics[width=\linewidth]{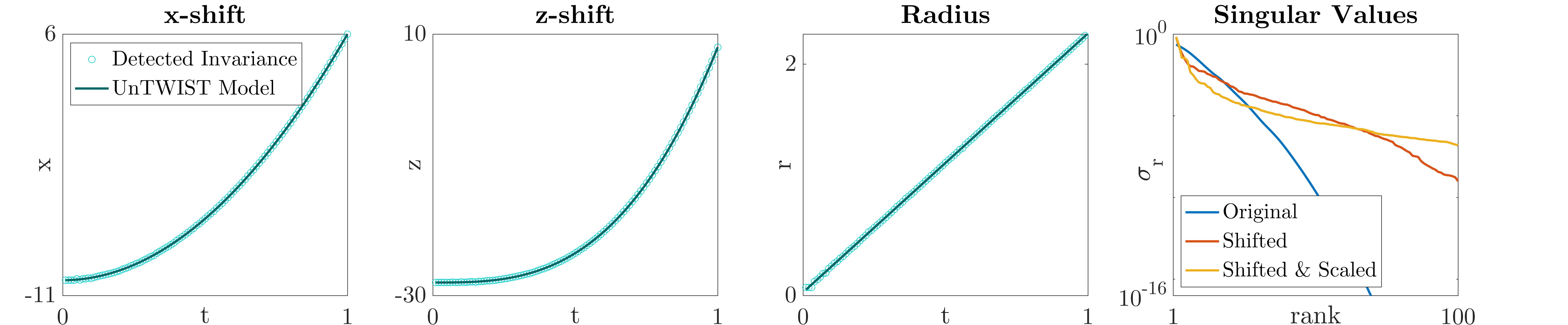} \par \medskip
	\includegraphics[width=\linewidth]{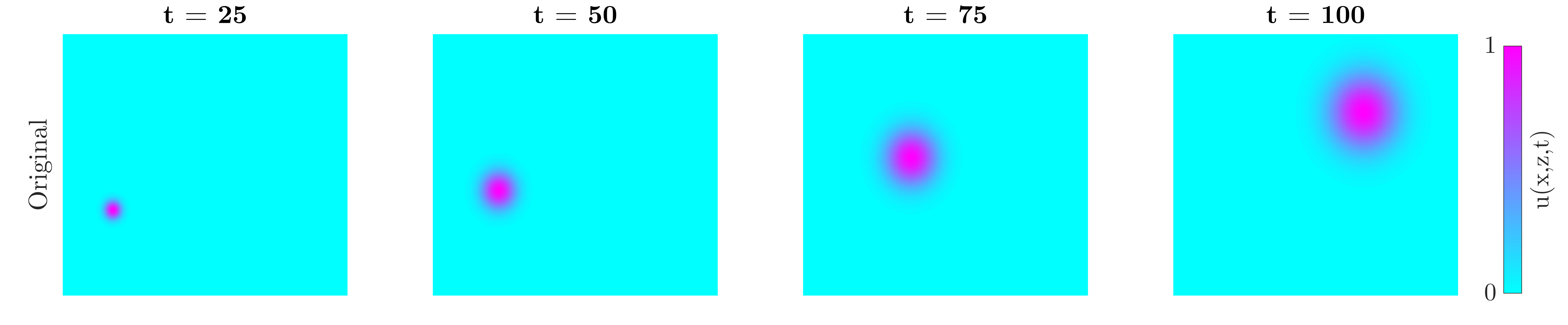}
	\includegraphics[width=\linewidth]{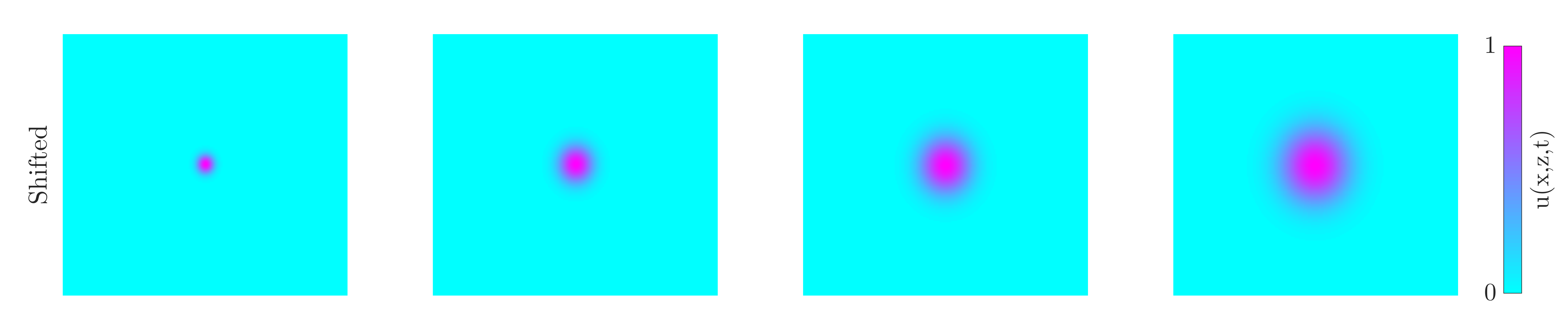}
	\includegraphics[width=\linewidth]{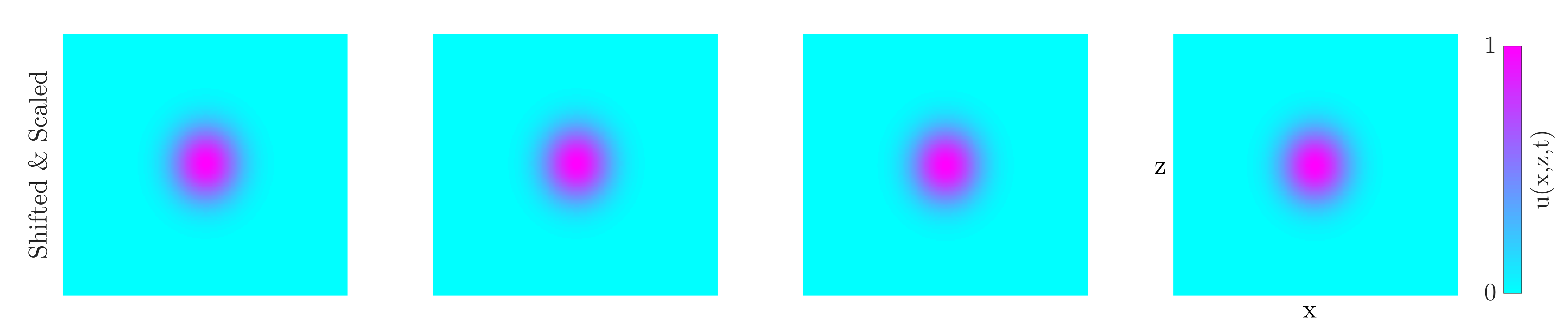}
	\caption[Shifting and scaling on 2D Gaussian example.]{(Top) Models discovered by the UnTWIST algorithm plotted alongside the detected invariances and singular value spectra of each level of invariance alignment. (Second) Original data, (Third) shifted data aligning the quantity of interest in the center of the frame. (Bottom) Effect of scaling based on the discovered radius model, yielding a perfectly stationary Gaussian.} 
	\label{fig:gauss_shiftscale}
\end{figure}

Using the initialization detailed in Section \ref{subsec:initialization}, shifts $\text{shift}_x $ and $\text{shift}_z$, and scaling parameter based on the radius $\mathbf{r}$ were determined for this data set. The UnTWIST algorithm was initialized with parameters $\lambda = 1$ and $\zeta = 1e-2$, and the following candidate speed library was used:
\begin{equation}
 \T(\mathbf{t}) =
 	\begin{bmatrix}
 	 \frac{1}{\mathbf{t}} & \mathbf{t}^4 & \mathbf{t}^3 & \mathbf{t}^2 & \sqrt{|\mathbf{t}|} & \mathbf{t} & \mathbf{1}
 	\end{bmatrix}.
\end{equation}
After 2 iterations, UnTWIST yields the following models for the invariances depicted in Figure~\ref{fig:gauss_shiftscale}.
\begin{equation}\label{eq:madeup_results}
\begin{split}
	\text{shift}_x(t) &= 4.0008 t ^2 - 2.0004 \\
	\text{shift}_z(t) &= 6.0007 t ^3 - 4.0003 \\
	s(t) &= 1.5095 t. 
\end{split}
\end{equation}
Compared to the known speed models given by~\eqref{eq:2dspeeds} and the true radial expansion given by~\eqref{eq:trueradius}, the resulting UnTWIST models perform well qualitatively and quantitatively, with a $\ell_2$-norm fit error of 6.7e-04. Of unique importance is that the models are sparse, match the correct form of the model terms, and are accurate in coefficient values, with a $\ell_2$-norm coefficient error of 1.3e-03. This validates the generalization of the UnTWIST algorithm in higher spatial dimensions and with a scaling invariance.

\begin{figure}[t!]
	\centering
	\includegraphics[width=\linewidth]{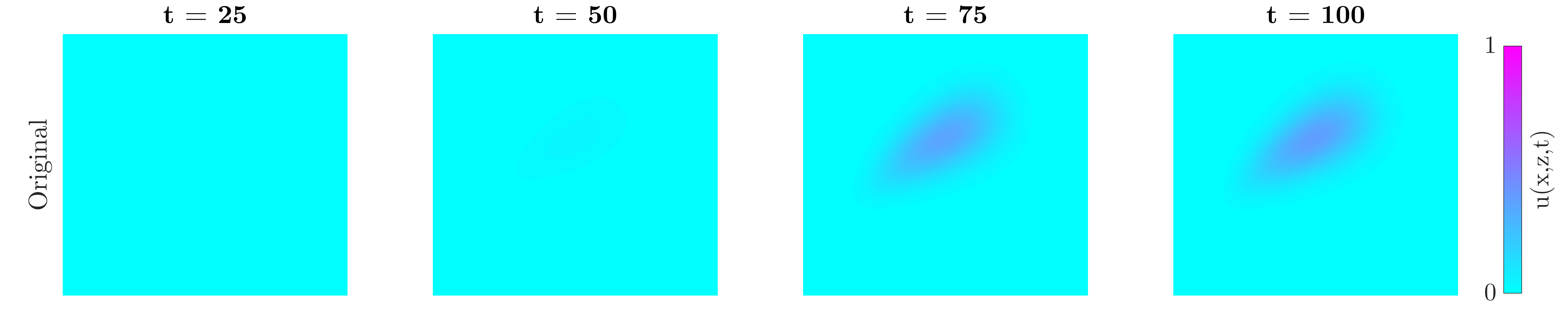}
	\includegraphics[width=\linewidth]{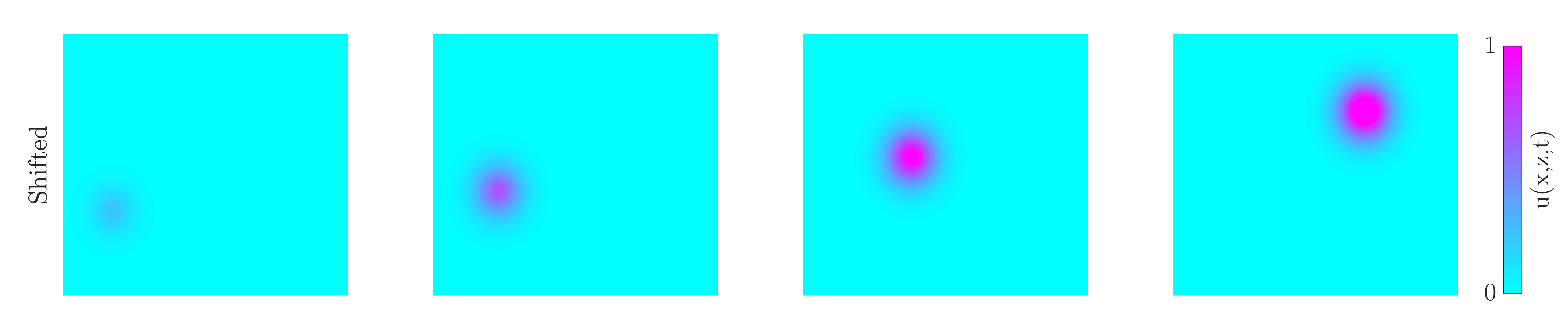}
	\includegraphics[width=\linewidth]{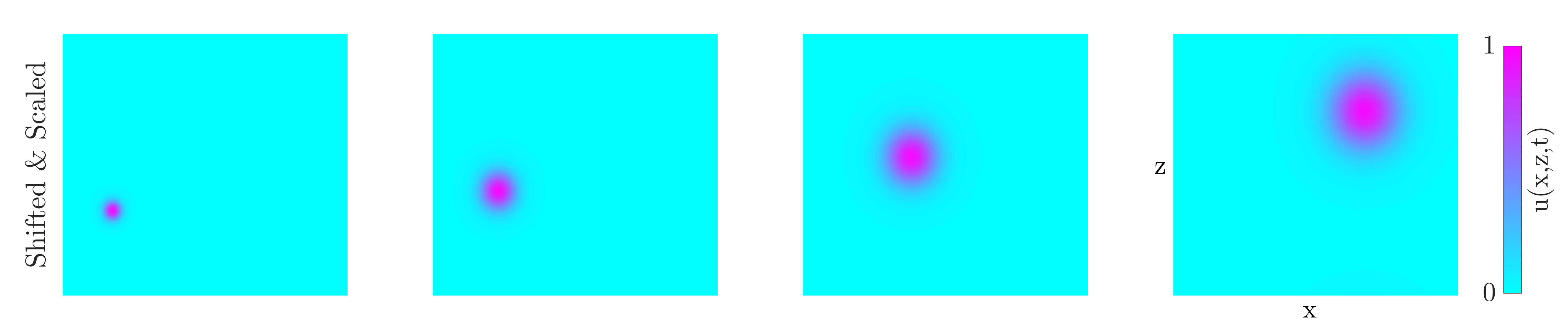}
	\caption[1-mode reconstructions of 2D Gaussian data]{Comparison of 1-mode reconstructions for each of the 3 stages of invariance reversal-- unadulterated original data, shifted data, and shifted \& scaled data. It is clear that 1 mode is insufficient for reconstructing the wave shapes for the original data, yields some information for the shifted data without capturing the scaling, and clearly describes the shifted \& scaled data. } 
	\label{fig:gauss_1mode}
\end{figure}

Once the models have been determined, the shifting and scaling operators can be applied. The results of these manipulations can be seen in Figure \ref{fig:gauss_shiftscale}. It is clear that applying the shifting and scaling operators makes the wave appear as a stationary quantity, which is more amenable to low-rank representation.  Specifically, the shifting and scaling invariances reduce the dynamics to an approximately rank-1 model. This is shown from by the singular value spectrum of each level of manipulation which discloses the role that these shifts play. The POD/SVD was computed for each data set and the decay of the singular values can be seen in Figure \ref{fig:gauss_shiftscale}. The shifted and scaled data's singular value decay shows a large share of the system's energy contained by the first mode, as well as the rapid drop off of the energy contained in the remaining modes. Relative to the original and shifted data, this indicates that very low rank representations may be suitable for this data. Reconstructions were generated for each data set. For the original data set, the first mode of the POD/SVD was retained. For the shifted data set, the first mode was retained, and the shift operation was inverted on this low-rank reconstruction. Similarly, for the shifted and scaled data set, the scaling was also inverted on the low-rank reconstruction, with zeros padding the matrix to account for down sampling. The comparison of these 1-mode reconstructions can be seen in Figure~\ref{fig:gauss_1mode}. It is clear from this comparison that for equivalently low-rank approximations, the shifting and scaling operators yield more physically relevant and interpretable reconstructions. Early in the original data's approximation, wave forms are not retained, with the spherical nature of the true wave being distorted. Conversely, the spherical wave shape, complete with traveling and scaling phenomena, is accurately captured when shifting and scaling are accounted for. 

Overall, the models discovered by UnTWIST prove to be physically relevant, accurate to the original system, and useful to improve the dimensionality reduction in both efficiency and interpretability. These promising results on this example problem indicate that the UnTWIST algorithm can be expected to perform similarly on systems with two-dimensional shifting and scaling phenomena. 

\subsection{One Detonation Wave}
\begin{figure}[t!]
    \centering
    \begin{subfigure}{\linewidth}
	\includegraphics[width = \linewidth]{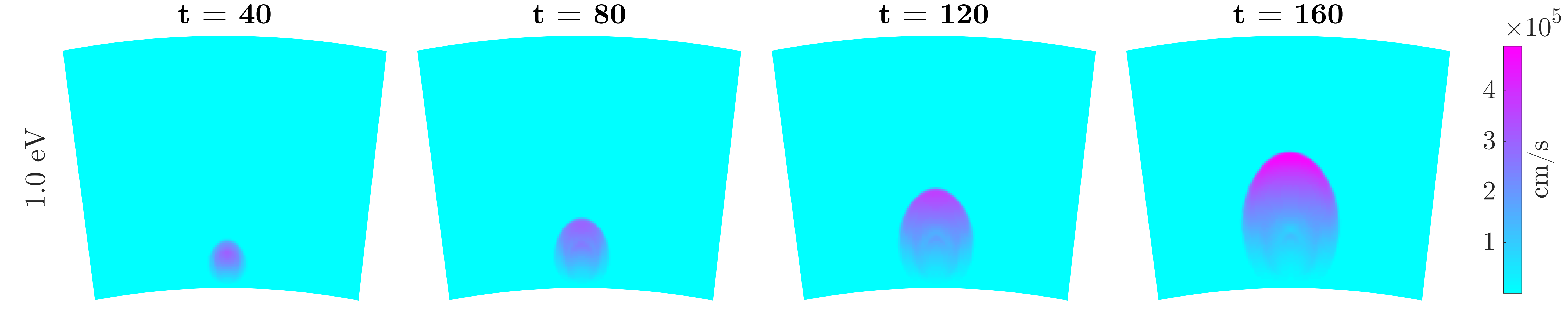}
    \end{subfigure}
    \begin{subfigure}{\linewidth}
	\includegraphics[width = \linewidth]{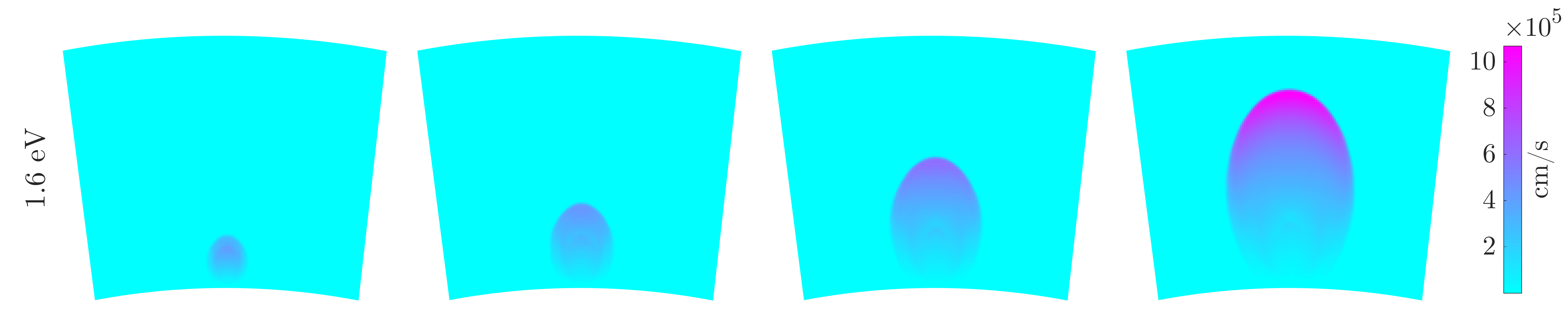}
    \end{subfigure}    \begin{subfigure}{\linewidth}
	\includegraphics[width = \linewidth]{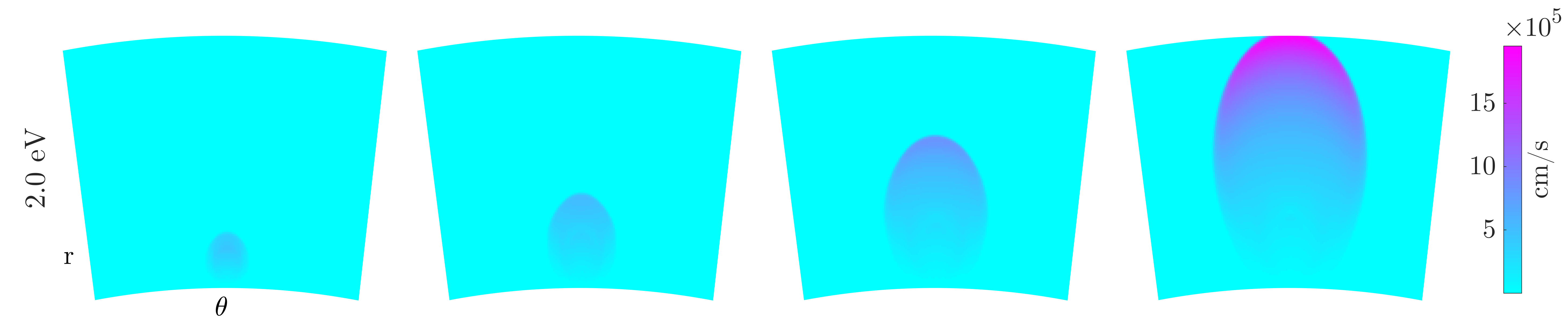}
    \end{subfigure}
    \caption[Time series of single detonations with various initial energies]{Velocity magnitude of $1.0$, $1.6$ and $2.0$ eV initial energy perturbation simulations at four example time slices as viewed from a central plane of the three-dimensional view.}
    \label{fig:initial_detonations}
\end{figure}

\paragraph{Data.} A 100 km radius hot sphere with no initial velocity is initialized in the center of the domain on the $\theta-\phi$ plane and at 200 km altitude. Three test cases with initial hot sphere energies of 1.0 eV, 1.6 eV and 2.0 eV were simulated. The resulting velocity magnitudes for three initial detonation energies can be seen in Figure~\ref{fig:initial_detonations}, depicting the center $\phi$ slice of the front at example times. Throughout the following examples, though three dimensions were computed, two-dimensional planar slices were considered. 

\begin{figure}[t!]
	\centering
		\includegraphics[width = \linewidth]{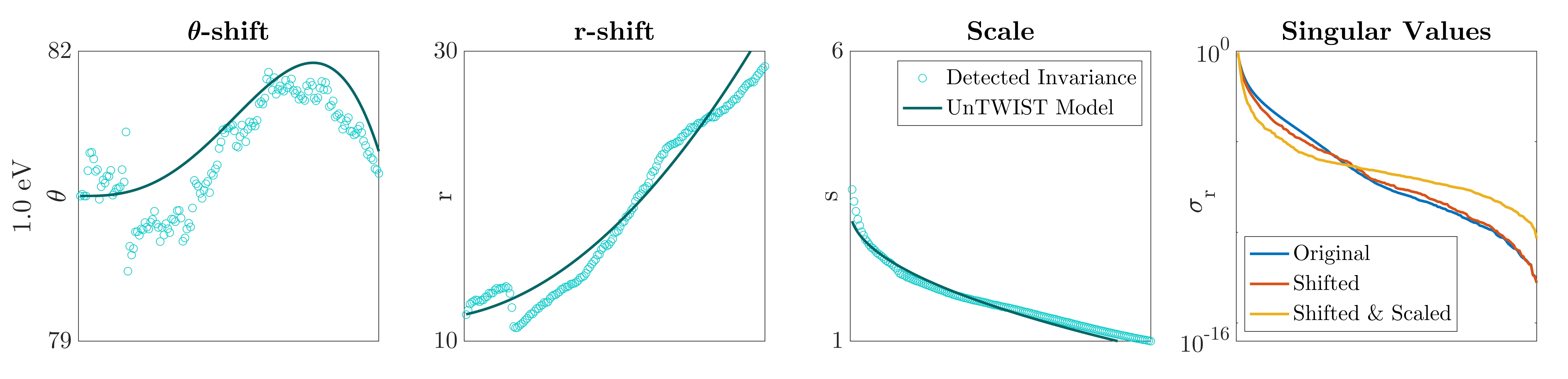}
		\includegraphics[width = \linewidth]{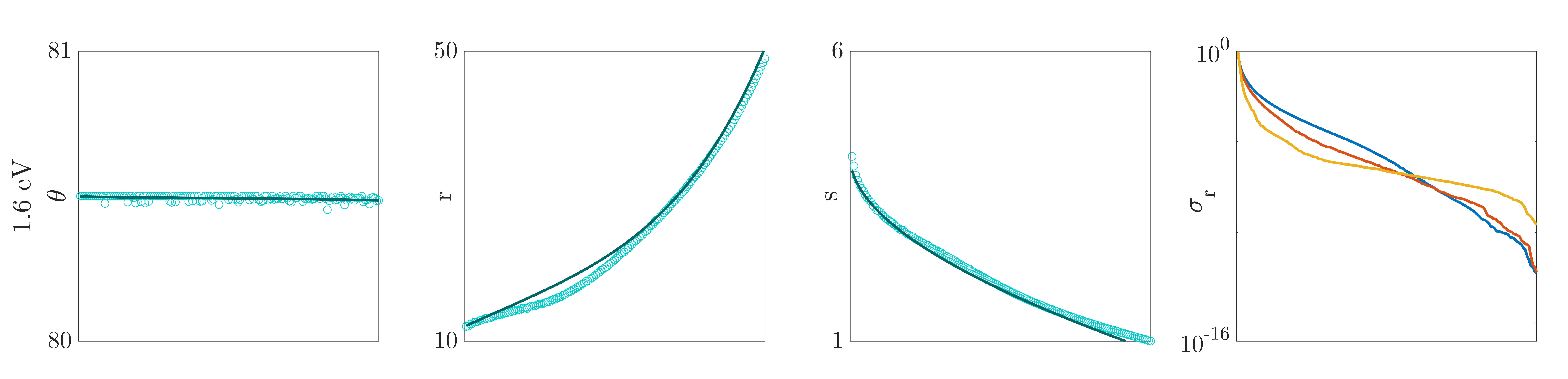}
		\includegraphics[width = \linewidth]{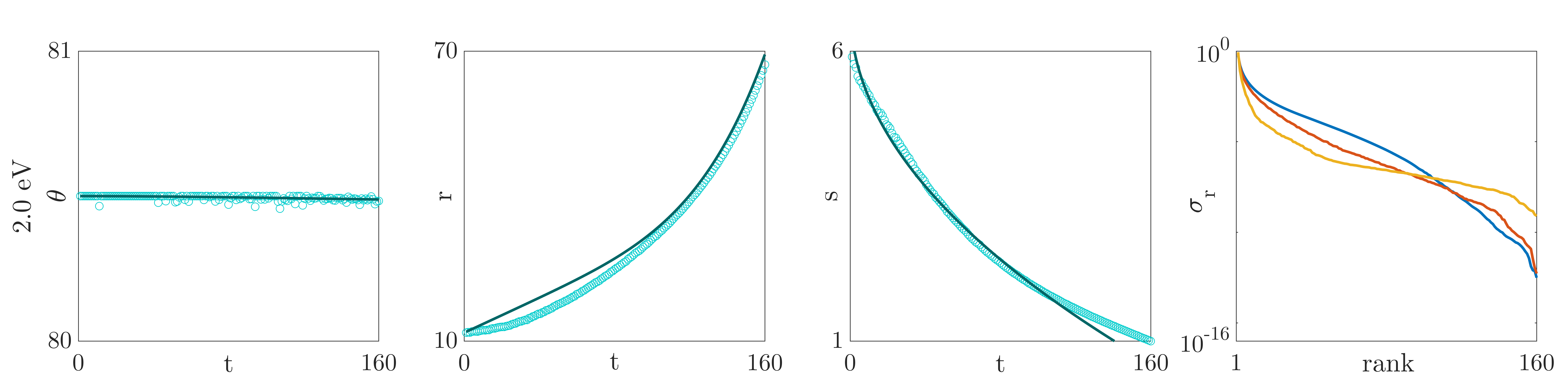}
	\caption[UnTWIST models and singular values for symmetries in each single wave simulation]{Models for shift$_\theta$, shift$_r$ and $\mathbf{s}(t)$ resulting from the UnTWIST algorithm for each initial perturbation energy, 1.0, 1.6, and 2.0 eV, compared to the data collected in the initialization step. Corresponding singular value spectra for each data set is also shown before and after accommodating for each invariance.}
	\label{fig:1wavemodels}
\end{figure}

\paragraph{UnTWIST Results.} This data, similar to the 2D Gaussian example, depicts three invariances: the rising of the spherical perturbation due to buoyancy, the potential drifting across the frame, and the increasing size.
The same steps shown in the previous example are used to process the Euler equation data, yielding $\theta$ and $r$ centers of the waves and radii over time. First, the UnTWIST algorithm was used to discover models for the $\theta$ and $r$ shifts and a scaling parameter for each data set. In this application, all values, including $t$, shift$_\theta$, shift$_r$, and $\mathbf{s}(t)$ were first normalized. In doing so, each model could be fit agnostic to the scale relative to the other quantities, i.e. $\mathbf{s}(t)$ fitting on the order of $10^1$ and shift$_r$ fitting on the order of $10^2$. After the model discovery step, the normalizations were accounted for. The identified wave centers, scaling parameter, as well as the discovered UnTWIST models for each quantity can be seen compared in Figure \ref{fig:1wavemodels} for each initial energy value. The input hyper-parameters, as well as the total iterations, model fit errors, and resulting model equations are given in Appendix A, Table~\ref{tab:1waveinfo}. 

Figure \ref{fig:1wavemodels} depicts the models for each data set for the three invariance we wish to account for-- the $\theta$-shift, $r$-shift, and scaling. While the data taken from initialization is somewhat noisy, especially as seen in the $\theta$- and $r$-shifts for the 1.0 eV detonation, it is clear that the UnTWIST models fit the data well while smoothing inconsistencies in the wave detection. This will be useful when accounting for these invariances in the data, yielding smooth transitions between time steps. 

These three invariances were accounted for in each data set, and a standard POD/SVD was computed for each data set. The resulting SVD can also be seen in Figure \ref{fig:1wavemodels}, which shows that the shifted and scaled data is most amenable to POD. The energy contained in the first mode is higher in the shifted and scaled data sets than the shifted or original data for each data set. The significance of the first few modes is most prominent in the 2.0 eV data set. In this case, the higher-energy wave front travels faster than the other initial energies. The improved initialization and model fit of UnTWIST on this data set indicates that performance of the UnTWIST method increases with wave speed. In addition, the singular value spectra indicate that very few modes are required to reconstruct all three data sets. 

\begin{figure}[t!]
	\centering
    \centering
    \begin{subfigure}{0.013\linewidth}
        \rotatebox{90}{\textbf{\footnotesize{1.0 eV}}}
    \end{subfigure}
    \begin{subfigure}{0.98\linewidth}
        \includegraphics[width =\linewidth]{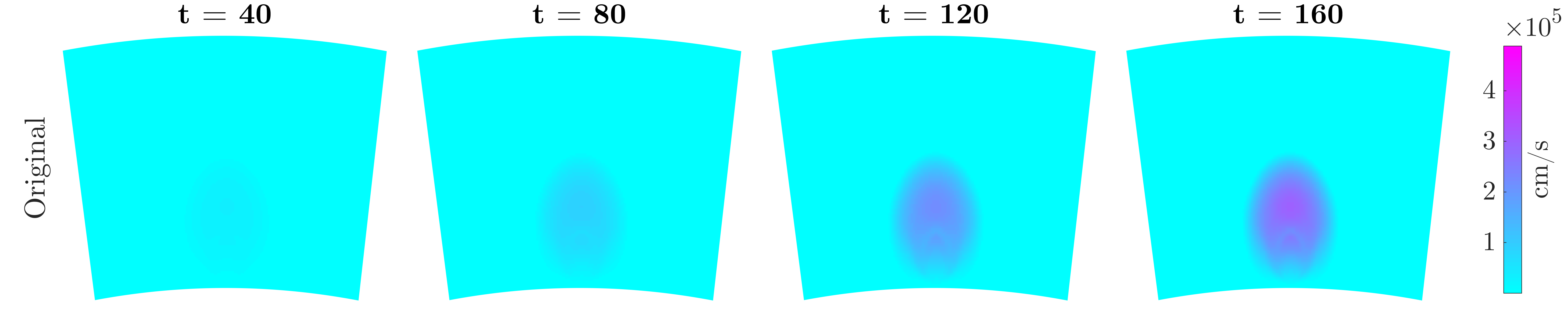}
        \includegraphics[width =\linewidth]{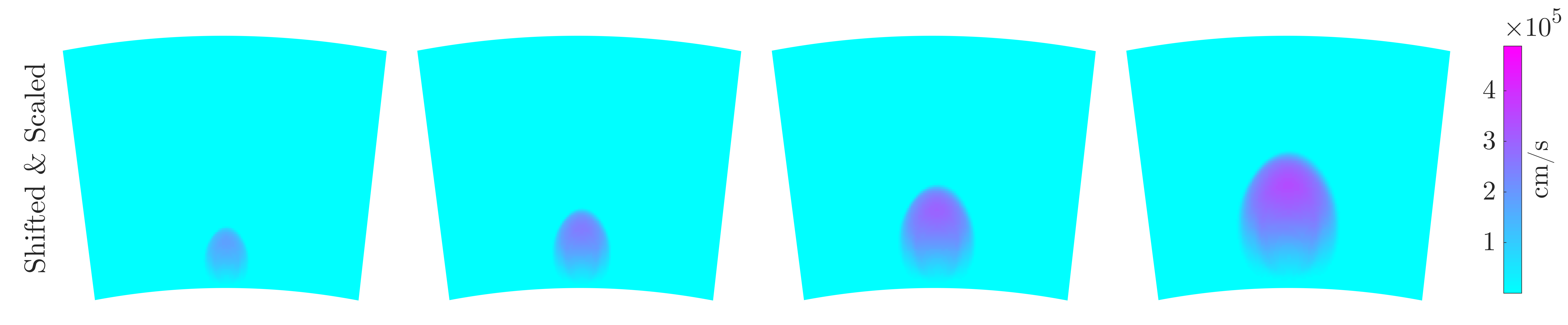}
	\end{subfigure}
	\begin{subfigure}{0.013\linewidth}
        \rotatebox{90}{\textbf{\footnotesize{1.6 eV}}}
    \end{subfigure}
    \begin{subfigure}{0.98\linewidth}
        \includegraphics[width =\linewidth]{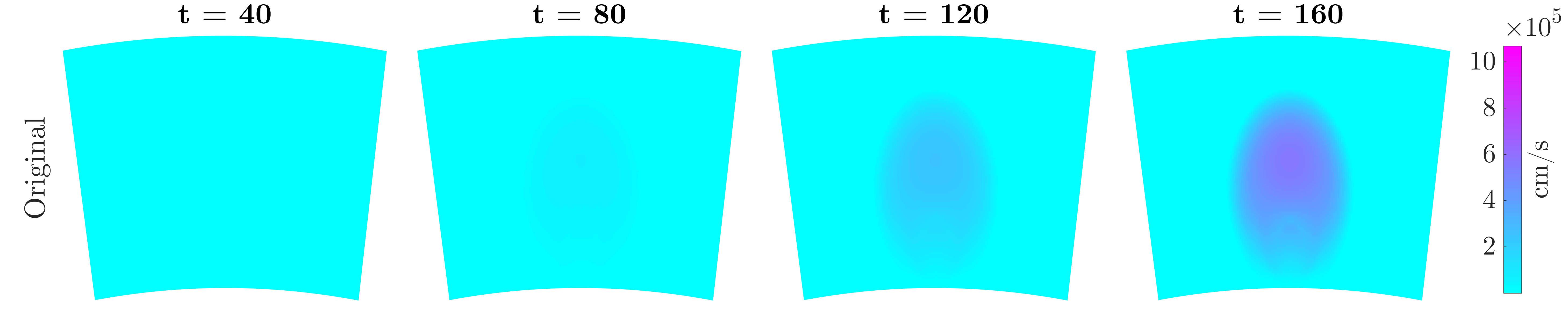}
        \includegraphics[width =\linewidth]{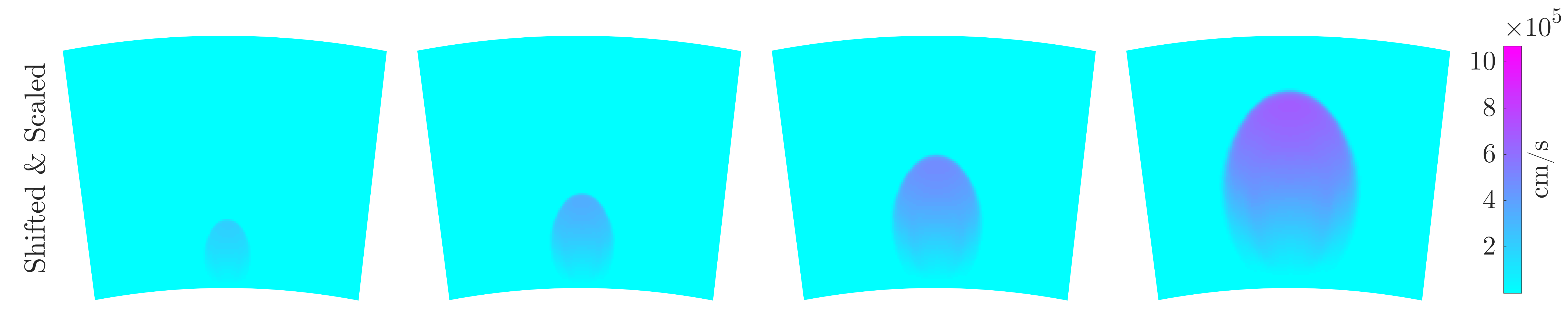}
	\end{subfigure}
	\begin{subfigure}{0.013\linewidth}
        \rotatebox{90}{\textbf{\footnotesize{2.0 eV}}}
    \end{subfigure}
    \begin{subfigure}{0.98\linewidth}
        \includegraphics[width =\linewidth]{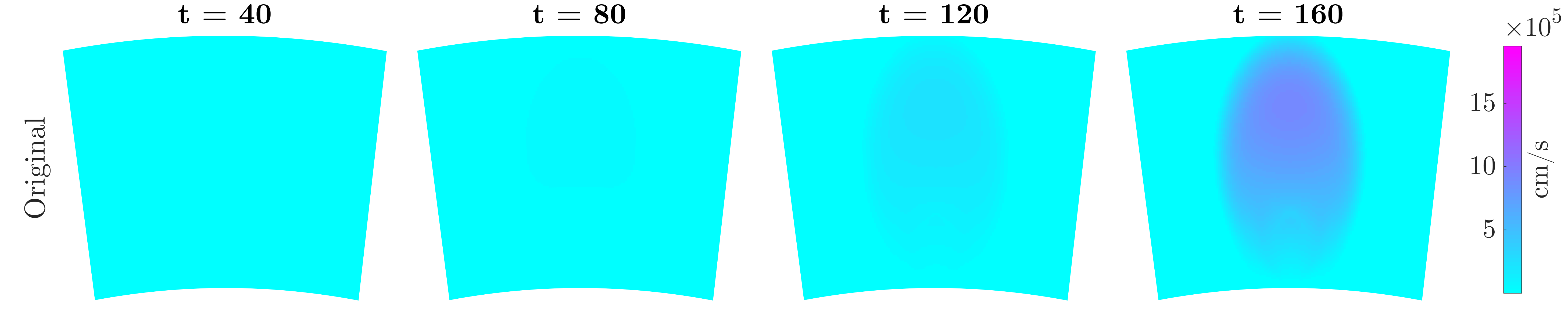}
        \includegraphics[width =\linewidth]{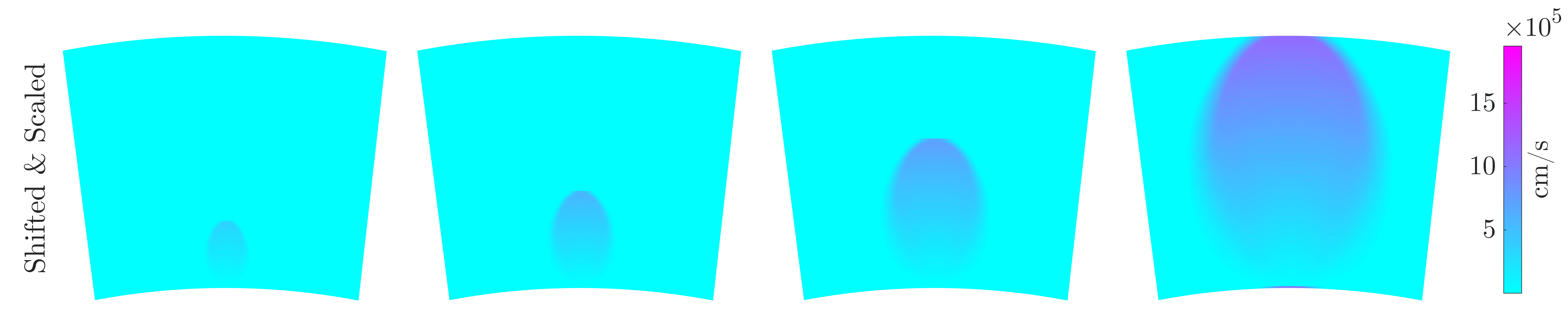}
	\end{subfigure}
	\caption{Rank-1 reconstructions of original vs UnTWIST-modified single-wave data.}
	\label{fig:1wave_reconstructions}
\end{figure}

For each data set, 1-POD-mode reconstructions were created, and invariance modifications were reversed as in the previous example. Figure~\ref{fig:1wave_reconstructions} shows these low-rank reconstructions. The results are overall effective and interpretable and show that UnTWIST has aided in the dimensionality reduction. For all data sets, the low-rank reconstruction of the original data set is unable to capture the invariances-- neither the translating nor the scaling is captured. For the 1.0 and 1.6 eV data sets, the UnTWIST-modified reconstruction is extremely close to the original data. Some of the intensity of the shock front is not captured toward the end of the time series, but the overall dynamics, as well as the shifting and scaling, are well captured. Similarly, the 2.0 eV data set captures these invariances, but presents some difficulties as the wave front exceeds the boundaries. In this case, the periodic shift adds the wave front to the bottom of the $r$-domain.  

\subsection{Two Initial Detonations}

While these results of a single detonation front are promising, one capability of UnTWIST has not yet been explored in this higher-dimensional example-- accommodating and separating multiple waves. 

\begin{figure}[tb!]
	\centering
	\includegraphics[width = \linewidth]{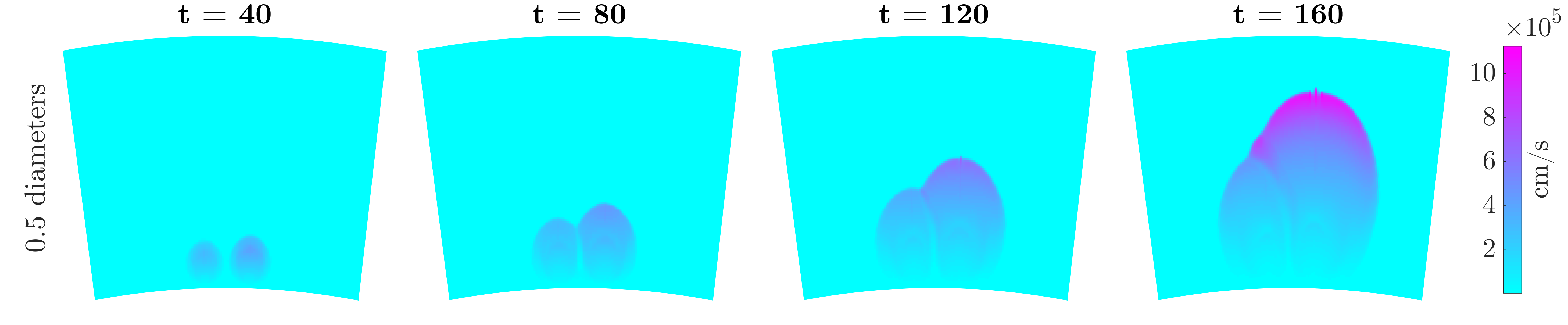}
	\includegraphics[width=\linewidth]{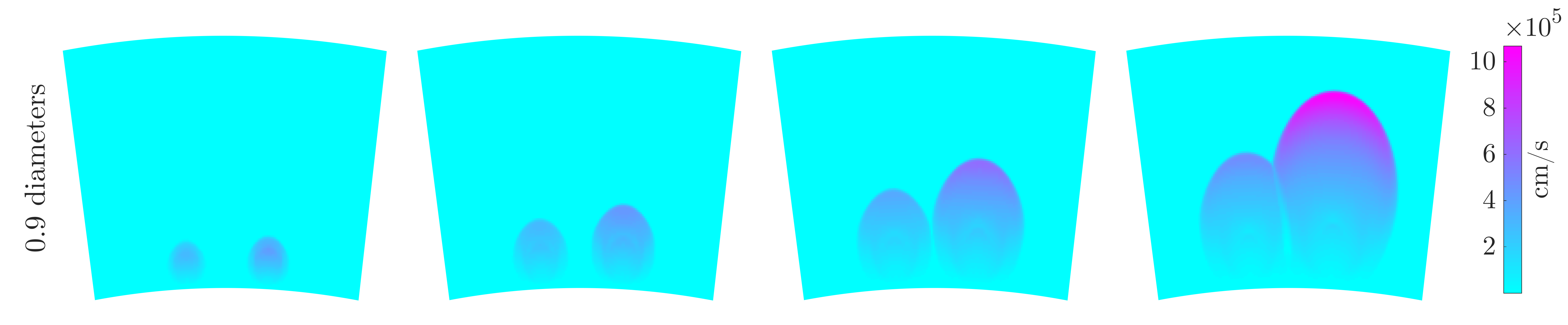}
	\includegraphics[width = \linewidth]{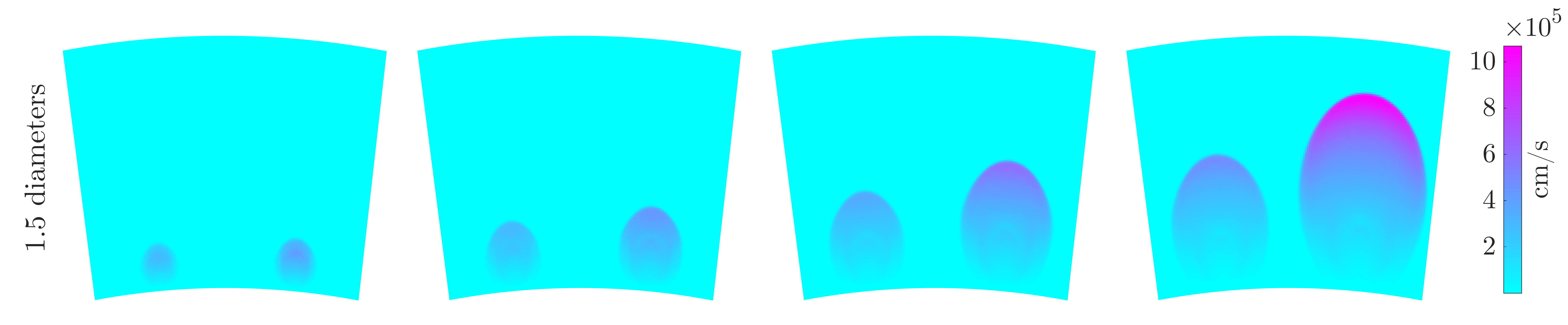}
\caption[Euler data sets with two waves]{Euler equation data with a 1.0 eV and 1.6 eV hot spheres initialized at various separations, 0.5-1.5 times the initial diameter. }\label{fig:2wavedata}
\end{figure}

\paragraph{Data.} Here, under the same domain, one hot sphere of 1.0 eV (left) and one of 1.6 eV (right) initial energies, each with a 100 km radius, are initialized. Three data sets are considered, with the spacing between the two detonations varying between 0.5 to 1.5 initial diameters apart. Time slices of this initial data can be seen in Figure \ref{fig:2wavedata}. As the initial separation distance between the wave fronts increases, the two quantities remain separate for a longer time before interfering. This interference will pose a challenge to UnTWIST, as the distinction between the wave becomes unclear, confusing the necessary invariance accommodation. 

\begin{figure}[tb!]
	\centering
	\includegraphics[width=\linewidth]{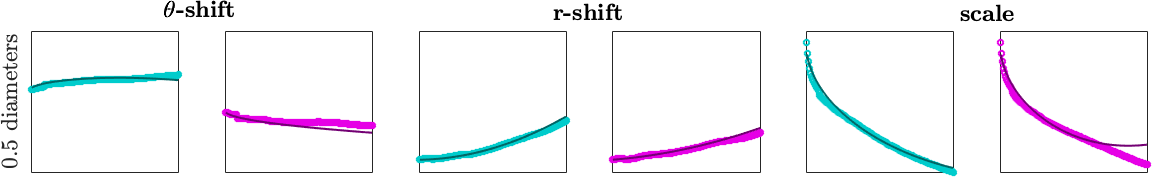} \par \medskip
	\includegraphics[width=\linewidth]{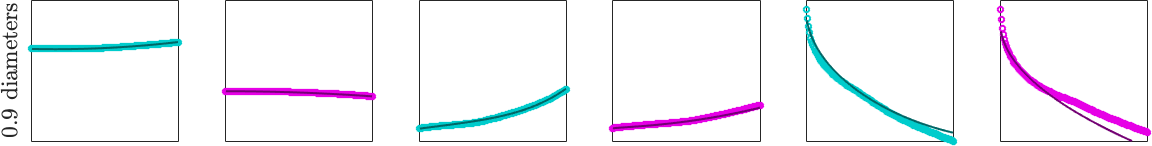} \par \medskip
    \includegraphics[width=\linewidth]{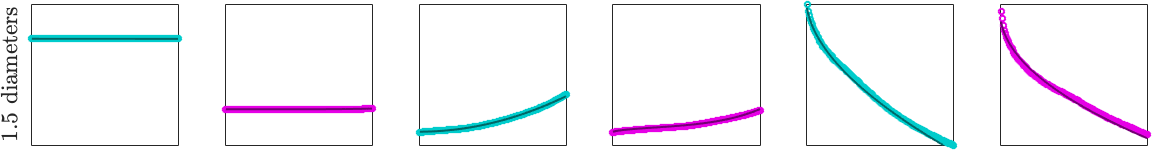}
\caption[UnTWIST models for two wave data]{UnTWIST models for shift$_\theta$, shift$_r$ and $\mathbf{s}(t)$ for wave 1 (teal) and wave 2 (pink) for the 0.5, 0.9, and 1.5 diameter initial detonation separation data sets.}\label{fig:2wavemodels}
\end{figure}

\paragraph{UnTWIST Results.}
Similar to the previous one-wave example, the same steps were performed for finding values for the wave centers and scalings. Here, the wave centers were naively computed by separating the domain into a left and right half. The same shift$_\theta$, shift$_r$ and $\mathbf{s}(t)$ parameters were computed for each data set. In this case, the separation function of UnTWIST is used, yielding models for wave 1 and wave 2 separately. The resulting models can be seen in Figure~\ref{fig:2wavemodels}. The details of the form of the models as well as the hyperparameters used in the optimization can be found in Appendix A. 

It is clear that from the increase in spacing comes more clear models for the invariances. Ideally, the $\theta$-shift for each data set would be captured as a constant value, since there is no apparent underlying translation in the $\theta$ direction. However, with close detonation wave fronts, such as the 0.5 diameter spacing case, the $\theta$-shift becomes muddled and veers toward the opposite wave. Figure~\ref{fig:2wavemodels} shows an improved picture of the $\theta$-shift with 1.5 diameter separation, with values for each wave fairly constant throughout time, with wave 1's $\theta$-shift being the ideal outcome. The intermediately spaced $\theta$-shift also shows that while the initial model may be accurate, the center of one individual wave is difficult to capture accurately when waves interact. The farther-spaced data set clearly shows the constant $\theta$-shift captured well until the waves collide and the models become skewed. 

The $r$-shifts are fairly accurately captured for both waves, though there is some noise in the two most closely-spaced examples. The scaling is well-captured, with all models matching closely to the data. This is in part due to the nature of the scaling computation, where each wave is assumed to be circular, yielding smooth radius and therefore scaling models. Nevertheless, the UnTWIST models capture well-fitting scaling parameters. 

\begin{figure}[tb!]
	\centering
	\includegraphics[width=0.32\linewidth]{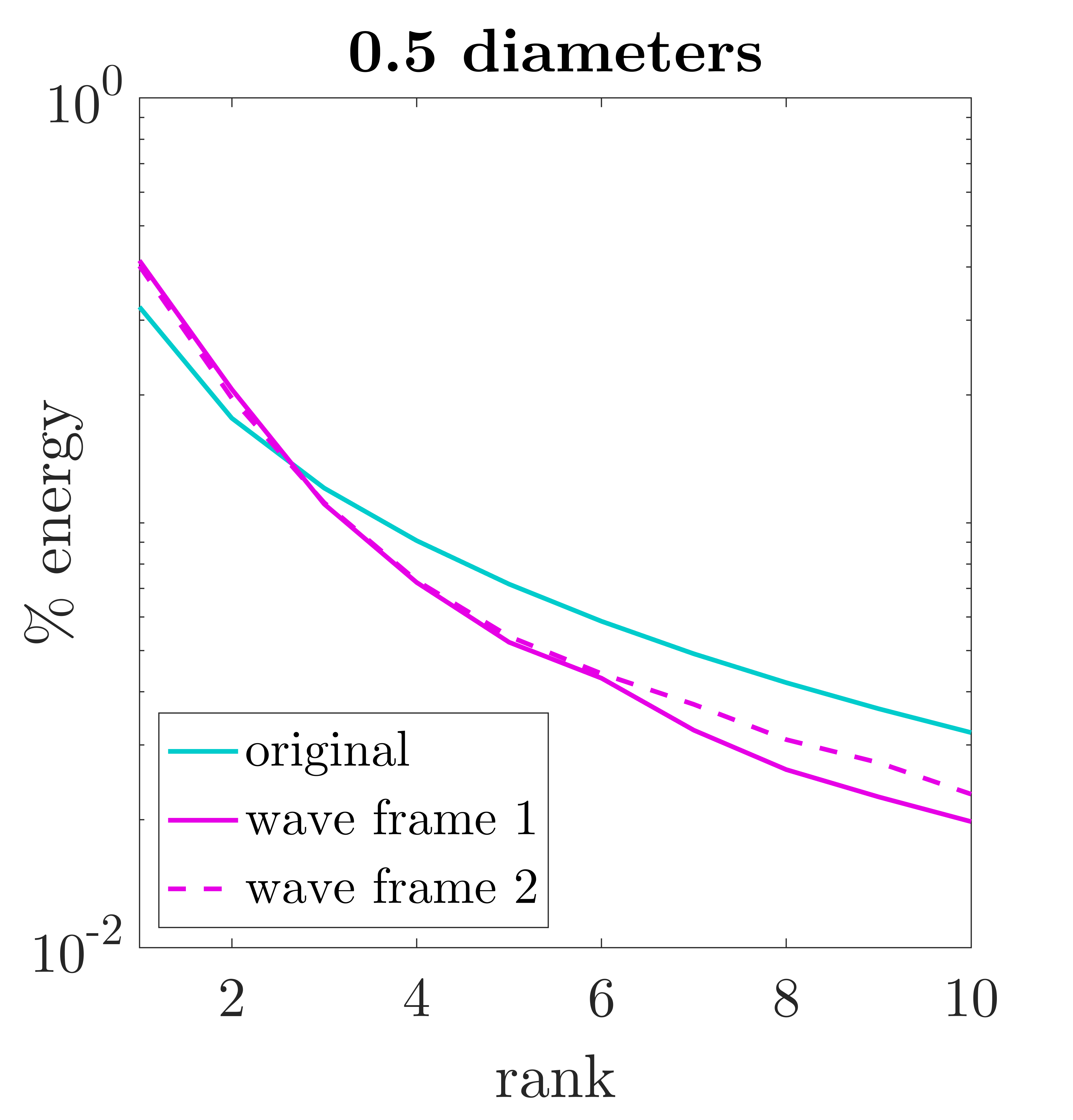}
	\includegraphics[width=0.32\linewidth]{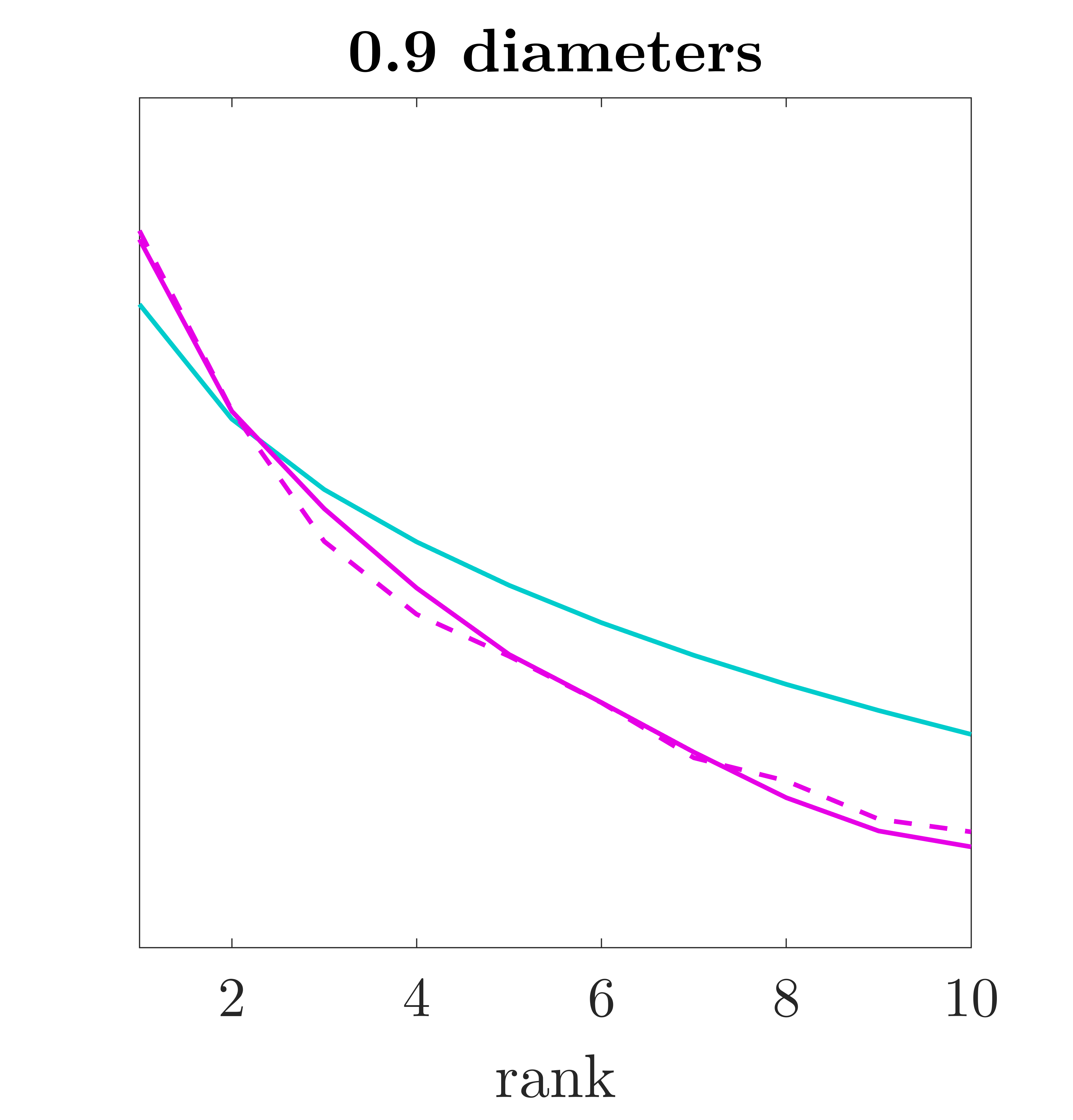}
	\includegraphics[width=0.32\linewidth]{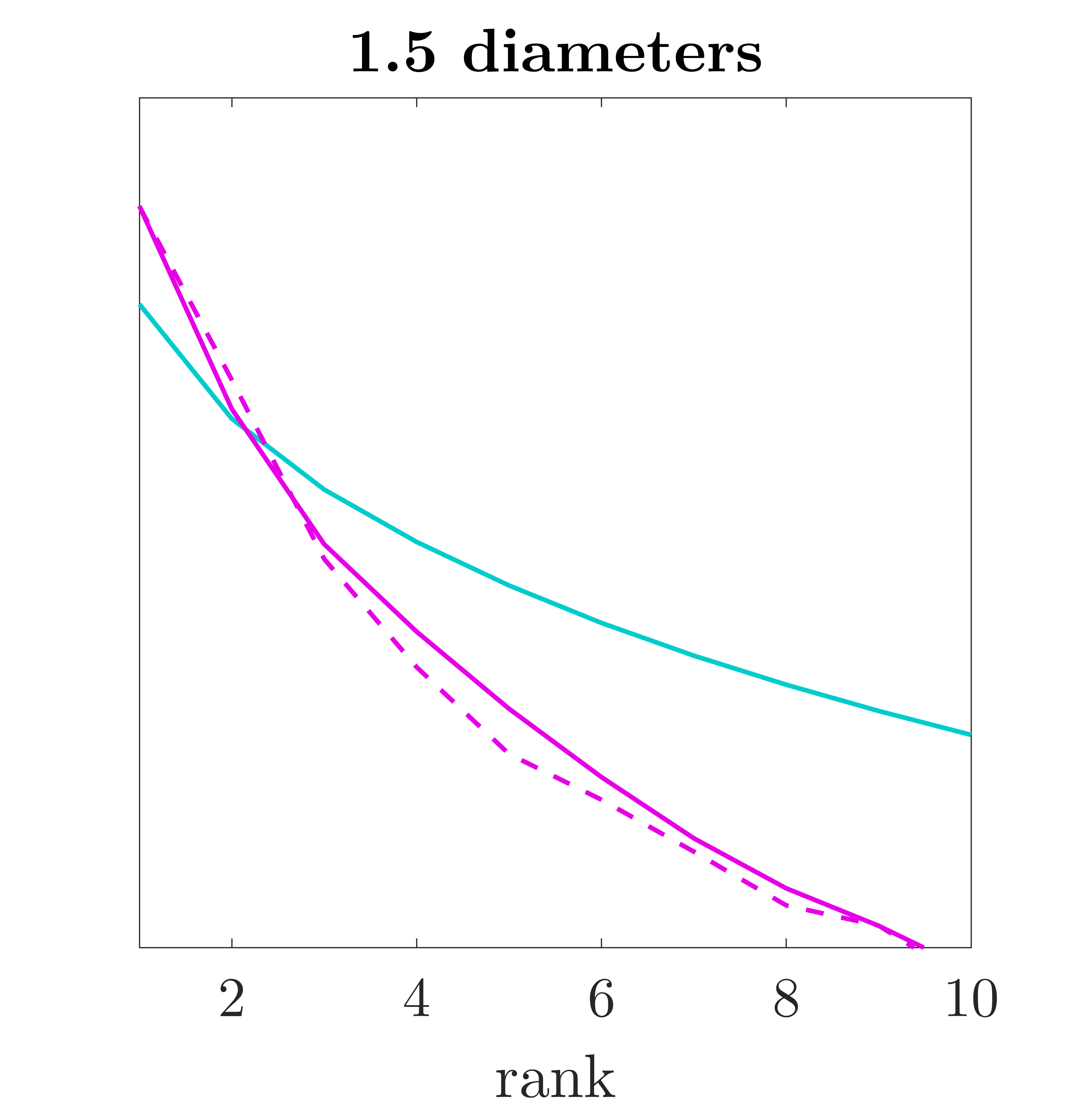}
	\caption[Singular value decompositions for two wave data]{Singular value decompositions of each two-wave data set. Blue lines shows the singular value decay of the original data set, where pink solid and dashed lines show the singular value decays after shifting \& scaling for each wave frame. }
	\label{fig:2wavesvd}
\end{figure}

In order to see the effect of these various invariance modifications, the shifting and scaling parameters were accounted for as in the previous examples. Whereas there are two desired wave frames, each aligned with respect to one wave, each wave frame was filtered using the predetermined threshold. Values outside the area of interest for that frame were set to zero. The entire data set was shifted and scaled to align each wave separately, and the singular value decomposition was taken for each wave frame. The results, compared to the singular value spectra of each original data set can be seen in Figure~\ref{fig:2wavesvd}.

\begin{figure}[tb!]
    \centering
    \begin{subfigure}{0.013\linewidth}
        \rotatebox{90}{\textbf{\footnotesize{0.5 diameters}}}
    \end{subfigure}
    \begin{subfigure}{0.98\linewidth}
        \includegraphics[width =\linewidth]{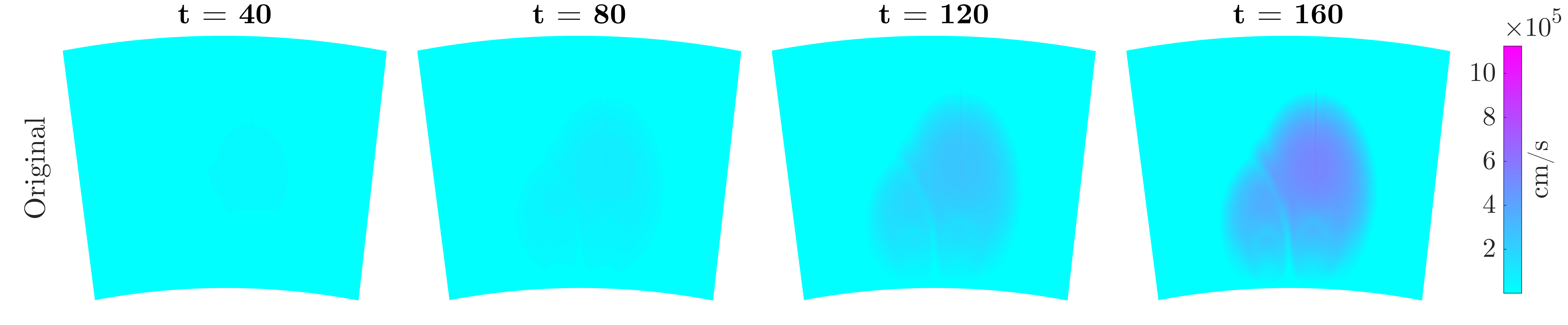}
	    \includegraphics[width =\linewidth]{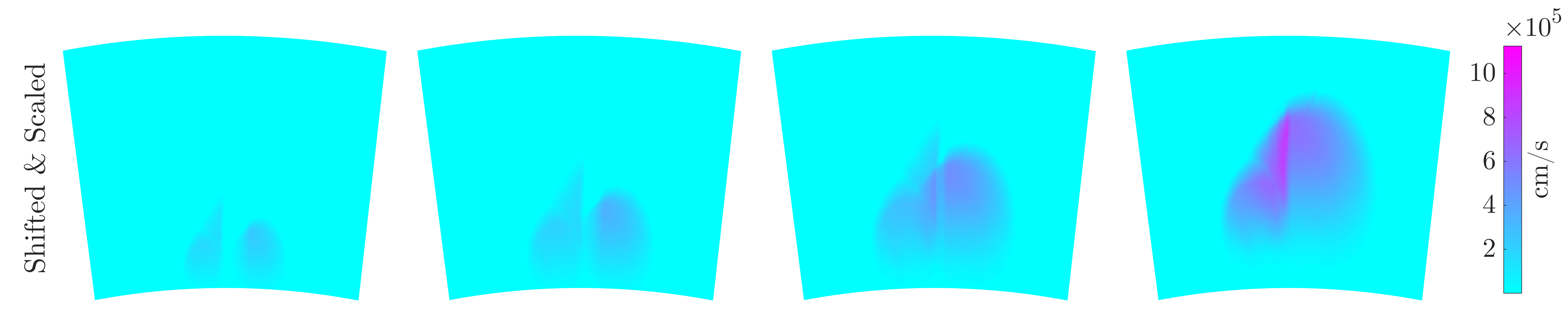}
	\end{subfigure}
	\begin{subfigure}{0.013\linewidth}
        \rotatebox{90}{\textbf{\footnotesize{0.9 diameters}}}
    \end{subfigure}
    \begin{subfigure}{0.98\linewidth}
        \includegraphics[width =\linewidth]{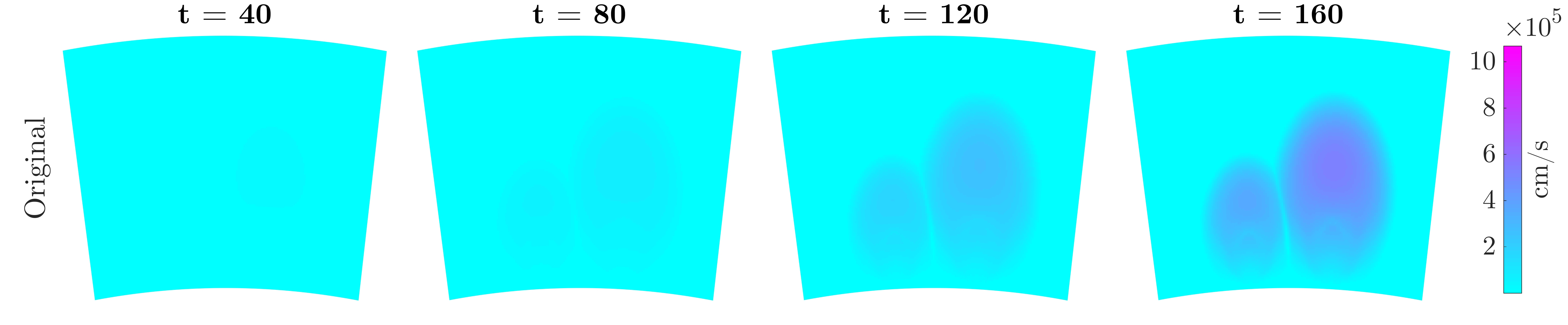}
	    \includegraphics[width =\linewidth]{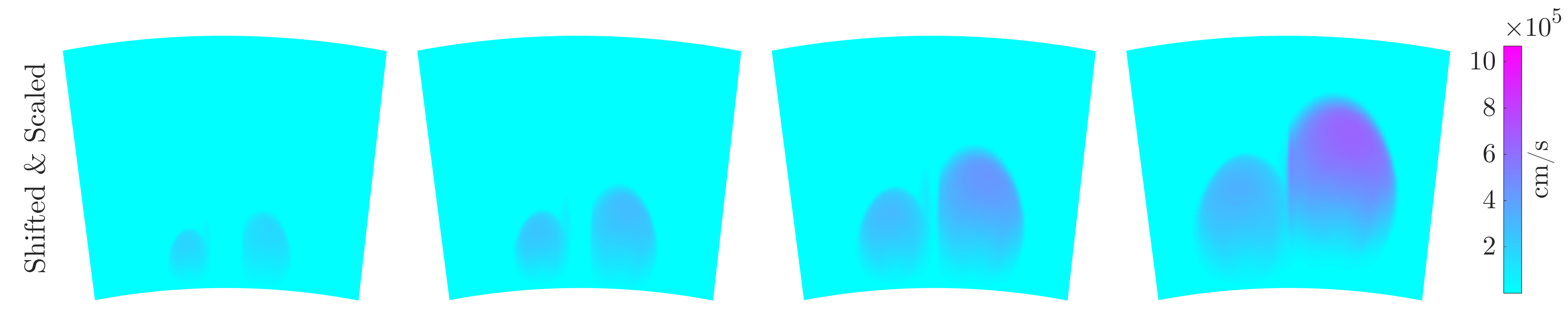}
	\end{subfigure}
	\begin{subfigure}{0.013\linewidth}
        \rotatebox{90}{\textbf{\footnotesize{1.5 diameters}}}
    \end{subfigure}
    \begin{subfigure}{0.98\linewidth}
        \includegraphics[width =\linewidth]{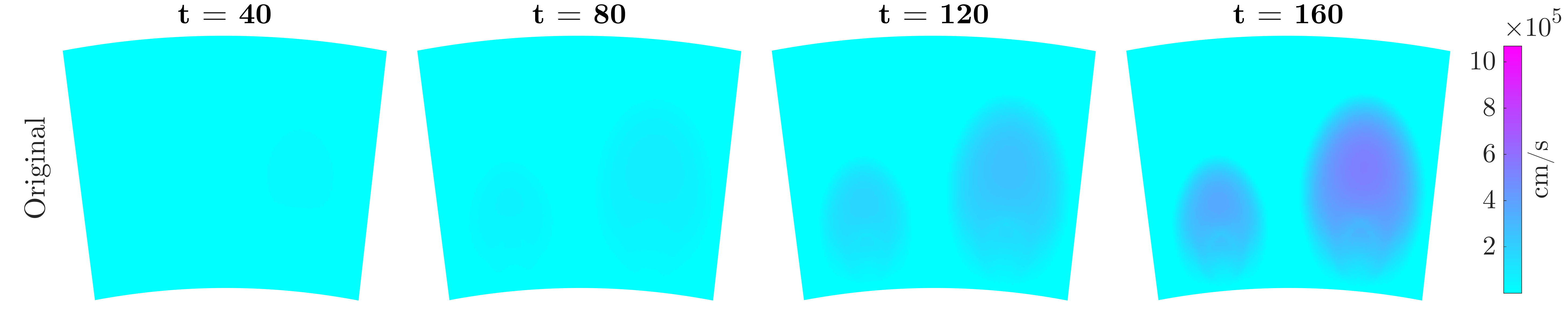}
	    \includegraphics[width =\linewidth]{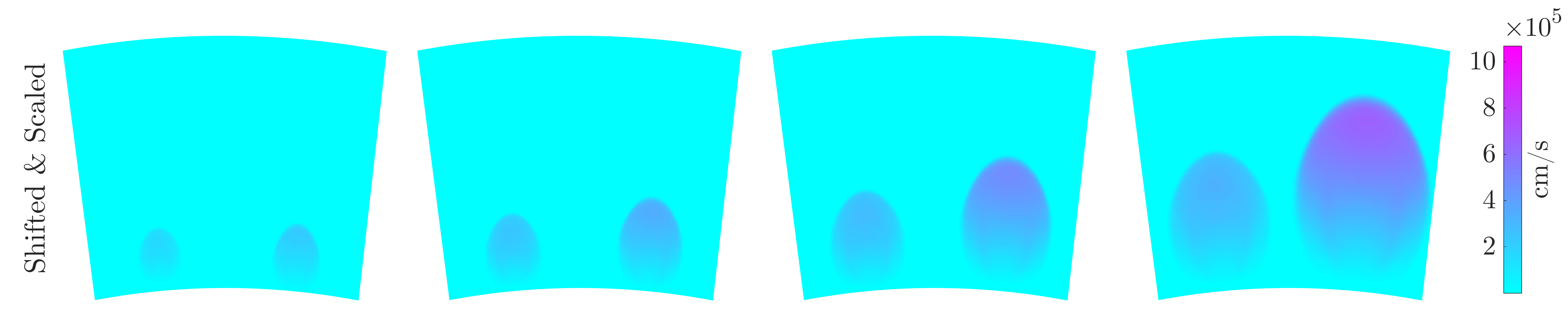}
	\end{subfigure}
	\caption{Rank-2 reconstructions of original vs UnTWIST-modified double-wave data.}
\label{fig:2waverecon}
\end{figure}

The compared singular value spectra in Figure~\ref{fig:2wavesvd} indicate that dimensionality reduction may be slightly improved. For many of the individual wave frames, the fraction of information captured by the first mode is slightly improved from the original data set. For some, the first mode performs noticeably better than the original data, but a closer look at the reconstructions is needed. To reconstruct the data using these low-rank modes, each wave frame was considered at a time. One mode was retained for each wave frame, and each wave frame was inversely shifted and scaled. The sum of the two low-rank reconstructions are used to represent the original data. The resulting 2-mode reconstructions for each data set can be seen in time slices in Figure~\ref{fig:2waverecon}, comparing the original data to the UnTWIST-modified data. 

From these reconstructions, UnTWIST shows a benefit over traditional dimensionality reduction approaches in cases where the waves are separable. For the farther-spaced data sets, such as the 1.5 diameter spacing, UnTWIST provides a clear advantage. The wave structures are accurately represented, retaining their separate scales and shifts. Compared to the reconstruction of the original data set, the fronts are clearly captured, and contain a larger share of the original energy. For more closely spaced wave fronts, the advantage of UnTWIST is less clear. Whereas the traditional POD approach does not capture the scaling accurately, and leaves out a large share of the energy in the center of the wave, the reconstructions using UnTWIST are also not physically realistic. In general, closely-spaced wave data sets may not be amenable to the current formulation due to the naive approach to separating the wave structures. Reconstructions exacerbate the hard boundaries used to detect invariance values. Alternatively, methods such as RPCA or other background subtraction approach~\cite{vaswani2018robust} may aid in separating these wave structures in their stationary wave frames. An improvement in the separation of the wave structures would allow UnTWIST to more viably represent these waves in low-rank. 

\section{Discussion and Conclusions}
In this work, the UnTWIST algorithm was generalized to accept inputs of  higher spatial dimensions with a broader class of invariances.  Specifically, two-dimensional data with scaling invariance were explored, with examples including a single-wave, noiseless system with known parameters, a single-wave system with more complex physics, and a two-wave system with various amounts of interaction between the two waves. In general, the UnTWIST algorithm was able to discover coordinates that allowed for improved, low-rank approximation ROMs for these invariances. In cases with close wave spacing, where individual wave quantities were difficult to parse, the UnTWIST algorithm was not able to accurately determine the invariances, and the reconstructions were poor. However, in moderately well-spaced example waves, low-rank reconstructions were improved, yielding more physically-relevant wave shapes while retaining the correct invariance phenomenon.  Importantly, the algorithmic improvements are modular, allowing for other invariances beyond translation and scaling considered here.  For instance, the mathematical infrastructure can allow for rotational invariance, for instance.  The focus on translation and scaling are related to the modeling of HANEs.  Specifically, the algorithms proposed can be used to structure improved ROMs for computationally tractable parametric studies of the effects of HANEs on atmospheric dynamics and global effects of such detonations.

There are still, however, clear improvements that can be made to the UnTWIST approach for two dimensional data with scaling. Primarily, the naive approach to invariance detection can be improved upon. In these examples, all wave centers were determined with a simple threshold. Accounting for quantities with different heights or intensities may be necessary for accurate wave detection. For example, identifying coherent vortical structures of various intensity would necessitate a more nuanced detection approach. Another suggested improvement is to the assumption that the wave quantities are approximately circular. With the scaling parameter dependent on approximate radial symmetry, severely non-symmetrical waves would be scaled incorrectly. Detection of the \textit{shape} of the wave quantity could be incorporated using more sophisticated image processing approaches or deep learning. Then, scaling in two dimensions separately could be utilized to account for the shape variations. Additionally, the scaling approach could be improved by utilizing a more sophisticated method for resizing data. Recent advances in super-resolution, specifically with deep learning~\cite{yang2019deep}, have made strides in accuracy and efficiency in super-resolving, or rescaling, single images. Overall, advanced object detection algorithms could be employed to better identify irregular wave quantities, with the UnTWIST algorithm applicable no matter the detection method. 

\section{Acknowledgments}
We acknowledge the support from the Defense Threat Reduction Agency HDTRA1-18-1-0038. AM acknowledges support from Graduate Opportunities and Minority Achievement Program Presidential Fellowship. 

\appendix \section{Appendix}
\begin{table}[h!]
	\centering
	\begin{tabular}{| c | c c c|} 
		\hline
		~ & \textbf{1.0 eV} & \textbf{1.6 eV} & \textbf{2.0 eV}\\
		\hline
		$\lambda$ & 5.0e-2 & 3e-1 & 1e-1\\
		$\zeta$ & 1e1 & 1e1 & 1e1\\
		iterations & 8,694 &  3,026 & 2,753\\
		shift$_\theta$ & $-11.1025t^4+11.5660t^3$ & $-0.0044t^5 -0.0105\sqrt{t}$ & $0.009t^3 -0.0206t^2$\\
		shift$_r$ & $15.3825t^2 + 4.4434t$ & $20.1387 t^4 + 18.2669 t$ & $ 28.6947t^5 + 28.8754 t$ \\
		$\mathbf{s}(t)$ & $-2.3937\sqrt{t} -0.3567$ & $-2.9889\sqrt{t} -0.3535t$ & $-6.1005\sqrt{t} + 0.8180$ \\ \hline
	\end{tabular}
	\caption[Hyperparameters, iterations, errors, and resulting models for single detonation simulations]{Hyperparameters, iterations, errors, and resulting models for Euler equation simulations of various initial energies.} \label{tab:1waveinfo}
\end{table}

	\begin{table}[h!]
		\centering
		\begin{tabular}{| c | c  c c|} 
			\hline
	        &	\textbf{0.5 diameters}	&	\textbf{0.9 diameters}	&	\textbf{1.5 diameters}	\\	\hline
$\lambda$	&	50e-1	    &	9e-1	    &	1e-1	\\	
$\zeta$	    &	1e1	        &	1e1	        &	1e1	\\	
iterations	&	4883	    &	4579	    &	1355	\\	
error	    &	1.949e-1	&	1.968e-1	&	3.929e-01	\\	
shift$_{\theta1}$	&	$-12.2t^2+22.7\sqrt{t}$	&	$11.6t^2-3.85t$	        &	$-0.00432t^5-0.00599\sqrt{t}$	\\	
shift$_{\theta2}$	&	$-25\sqrt{t}-24.1$	    &	$-5.83t^2-48$	        &	$0.791t^5-80$	\\	
shift$_{r1}$	&	$39.4t^2+2.32t$	        &	$20.4t^4+18.1t$	        &	$34.2t^2+0.807$	\\	
shift$_{r2}$	&	$19.3t^2+11.3t$	        &	$16.2t^2+4.04\sqrt{t}$	&	$13.7t^4+8.86\sqrt{t}$	\\	
$\mathbf{s}_1(t)$	&	$0.386t^2-3.05\sqrt{t}$	&	$0.379t^2-3.01\sqrt{t}$	&	$-2.95\sqrt{t}-0.333t$	\\	
$\mathbf{s}_2(t)$	&	$0.959t^2-3.12\sqrt{t}$	&	$-2.69\sqrt{t}-0.259$	&	$-2.7\sqrt{t}-0.313$	\\	\hline
		\end{tabular}
		\caption[Hyperparameters, iterations, errors, and resulting models for two-detonation simulations]{Hyperparameters, iterations, errors, and resulting models for two-detonation Euler equation simulations of various initial separation distances.} \label{tab:2waveinfo}
	\end{table}

\bibliography{main.bib}

\begin{thebibliography}{10}

\bibitem{Taira2017aiaa}
Kunihiko Taira, Steven~L Brunton, Scott Dawson, Clarence~W Rowley, Tim
  Colonius, Beverley~J McKeon, Oliver~T Schmidt, Stanislav Gordeyev, Vassilios
  Theofilis, and Lawrence~S Ukeiley.
\newblock Modal analysis of fluid flows: An overview.
\newblock {\em AIAA Journal}, 55(12):4013--4041, 2017.

\bibitem{taira2020modal}
Kunihiko Taira, Maziar~S Hemati, Steven~L Brunton, Yiyang Sun, Karthik
  Duraisamy, Shervin Bagheri, Scott~TM Dawson, and Chi-An Yeh.
\newblock Modal analysis of fluid flows: Applications and outlook.
\newblock {\em AIAA journal}, 58(3):998--1022, 2020.

\bibitem{benner2015survey}
Peter Benner, Serkan Gugercin, and Karen Willcox.
\newblock A survey of projection-based model reduction methods for parametric
  dynamical systems.
\newblock {\em SIAM review}, 57(4):483--531, 2015.

\bibitem{Kutz:2013}
J.~N. Kutz.
\newblock {\em Data-Driven Modeling \& Scientific Computation: Methods for
  Complex Systems \& Big Data}.
\newblock Oxford University Press, 2013.

\bibitem{Brunton2019book}
S.~L. Brunton and J.~N. Kutz.
\newblock {\em Data-Driven Science and Engineering: Machine Learning, Dynamical
  Systems, and Control}.
\newblock Cambridge University Press, 2019.

\bibitem{antoulas2005approximation}
Athanasios~C Antoulas.
\newblock {\em Approximation of large-scale dynamical systems}, volume~6.
\newblock Siam, 2005.

\bibitem{quarteroni2015reduced}
Alfio Quarteroni, Andrea Manzoni, and Federico Negri.
\newblock {\em Reduced basis methods for partial differential equations: an
  introduction}, volume~92.
\newblock Springer, 2015.

\bibitem{hesthaven2016certified}
Jan~S Hesthaven, Gianluigi Rozza, Benjamin Stamm, et~al.
\newblock {\em Certified reduced basis methods for parametrized partial
  differential equations}.
\newblock Springer, 2016.

\bibitem{Schmid2010jfm}
P.~J. Schmid.
\newblock Dynamic mode decomposition of numerical and experimental data.
\newblock {\em Journal of Fluid Mechanics}, 656:5--28, August 2010.

\bibitem{Rowley2009jfm}
C.~W. Rowley, I.\ Mezi\'c, S.\ Bagheri, P.\ Schlatter, and D.S. Henningson.
\newblock Spectral analysis of nonlinear flows.
\newblock {\em J.\ Fluid Mech.}, 645:115--127, 2009.

\bibitem{Tu2014jcd}
J.~H. Tu, C.~W. Rowley, D.~M. Luchtenburg, S.~L. Brunton, and J.~N. Kutz.
\newblock On dynamic mode decomposition: theory and applications.
\newblock {\em J. Comp. Dyn.}, 1(2):391--421, 2014.

\bibitem{Kutz2016book}
J.~N. Kutz, S.~L. Brunton, B.~W. Brunton, and J.~L. Proctor.
\newblock {\em Dynamic Mode Decomposition: Data-Driven Modeling of Complex
  Systems}.
\newblock SIAM, 2016.

\bibitem{Candes:2011}
E.~J. Cand\`es, X.~Li, Y.~Ma, and J.~Wright.
\newblock Robust principal component analysis?
\newblock {\em Journal of the ACM}, 58(3):11--1--11--37, 2011.

\bibitem{Towne2018jfm}
Aaron Towne, Oliver~T Schmidt, and Tim Colonius.
\newblock Spectral proper orthogonal decomposition and its relationship to
  dynamic mode decomposition and resolvent analysis.
\newblock {\em Journal of Fluid Mechanics}, 847:821--867, 2018.

\bibitem{willcox2006cf}
Karen Willcox.
\newblock Unsteady flow sensing and estimation via the gappy proper orthogonal
  decomposition.
\newblock {\em Computers \& fluids}, 35(2):208--226, 2006.

\bibitem{rowley2005model}
Clarence~W Rowley.
\newblock Model reduction for fluids, using balanced proper orthogonal
  decomposition.
\newblock {\em International Journal of Bifurcation and Chaos},
  15(03):997--1013, 2005.

\bibitem{berkooz1993proper}
Gal Berkooz, Philip Holmes, and John~L Lumley.
\newblock The proper orthogonal decomposition in the analysis of turbulent
  flows.
\newblock {\em Annual review of fluid mechanics}, 25(1):539--575, 1993.

\bibitem{kirby1992reconstructing}
Michael Kirby and Dieter Armbruster.
\newblock Reconstructing phase space from {PDE} simulations.
\newblock {\em Zeitschrift f{\"u}r angewandte Mathematik und Physik ZAMP},
  43(6):999--1022, 1992.

\bibitem{Rowley2000physd}
Clarence~W Rowley and Jerrold~E Marsden.
\newblock Reconstruction equations and the {K}arhunen--{L}o{\`e}ve expansion
  for systems with symmetry.
\newblock {\em Physica D: Nonlinear Phenomena}, 142(1-2):1--19, 2000.

\bibitem{Rim2018juq}
Donsub Rim, Scott Moe, and Randall~J LeVeque.
\newblock Transport reversal for model reduction of hyperbolic partial
  differential equations.
\newblock {\em SIAM/ASA J. Uncertainty Quantification}, 6(1):118--150, 2018.

\bibitem{Reiss2018jsc}
Julius Reiss, Philipp Schulze, J{\"o}rn Sesterhenn, and Volker Mehrmann.
\newblock The shifted proper orthogonal decomposition: A mode decomposition for
  multiple transport phenomena.
\newblock {\em SIAM Journal on Scientific Computing}, 40(3):A1322--A1344, 2018.

\bibitem{carlberg2015adaptive}
Kevin Carlberg.
\newblock Adaptive h-refinement for reduced-order models.
\newblock {\em International Journal for Numerical Methods in Engineering},
  102(5):1192--1210, 2015.

\bibitem{mendible2020dimensionality}
Ariana Mendible, Steven~L Brunton, Wes Lowrie, Aleksander~Y Aravkin, and
  J~Nathan Kutz.
\newblock Dimensionality reduction and reduced-order modeling for traveling
  wave physics.
\newblock {\em Theor. Comput. Fluid Dyn}, 2020.

\bibitem{mendible2021}
Ariana Mendible.
\newblock Data-driven modeling of rotating detonation waves.
\newblock {\em Physical Review Fluids}, 6(5), 2021.

\bibitem{hess1964effects}
Wilmot~N Hess.
\newblock {\em The effects of high altitude explosions}.
\newblock Citeseer, 1964.

\bibitem{hoerlin1976united}
Herman Hoerlin.
\newblock United states high-altitude test experiences.
\newblock {\em Rep. LA-6J05, Los Alamos Natl. Lab., Los Alamos, NM}, 1976.

\bibitem{arendt1964effects}
PR~Arendt and H~Soicher.
\newblock Effects of arctic nuclear explosions on satellite radio
  communication.
\newblock {\em Proceedings of the IEEE}, 52(6):672--676, 1964.

\bibitem{clawpack}
{Clawpack Development Team}.
\newblock Clawpack software, 2020.
\newblock Version 5.7.0.

\bibitem{mandli2016clawpack}
Kyle~T Mandli, Aron~J Ahmadia, Marsha Berger, Donna Calhoun, David~L George,
  Yiannis Hadjimichael, David~I Ketcheson, Grady~I Lemoine, and Randall~J
  LeVeque.
\newblock Clawpack: building an open source ecosystem for solving hyperbolic
  pdes.
\newblock {\em PeerJ Computer Science}, 2:e68, 2016.

\bibitem{pyclaw-sisc}
David~I. Ketcheson, Kyle~T. Mandli, Aron~J. Ahmadia, Amal Alghamdi, Manuel
  {Quezada de Luna}, Matteo Parsani, Matthew~G. Knepley, and Matthew Emmett.
\newblock {PyClaw: Accessible, Extensible, Scalable Tools for Wave Propagation
  Problems}.
\newblock {\em SIAM Journal on Scientific Computing}, 34(4):C210--C231,
  November 2012.

\bibitem{BaleLevMitRoss02}
D.~Bale, R.~J. LeVeque, S.~Mitran, and J.~A. Rossmanith.
\newblock A wave-propagation method for conservation laws and balance laws with
  spatially varying flux functions.
\newblock {\em SIAM J. Sci. Comput.}, 24:955--978, 2002.

\bibitem{ng2002spectral}
Andrew~Y Ng, Michael~I Jordan, and Yair Weiss.
\newblock On spectral clustering: Analysis and an algorithm.
\newblock In {\em Advances in neural information processing systems}, pages
  849--856, 2002.

\bibitem{zheng2018unified}
Peng Zheng, Travis Askham, Steven~L Brunton, J~Nathan Kutz, and Aleksandr~Y
  Aravkin.
\newblock A unified framework for sparse relaxed regularized regression: Sr3.
\newblock {\em IEEE Access}, 7:1404--1423, 2018.

\bibitem{vaswani2018robust}
Namrata Vaswani, Thierry Bouwmans, Sajid Javed, and Praneeth Narayanamurthy.
\newblock Robust subspace learning: Robust pca, robust subspace tracking, and
  robust subspace recovery.
\newblock {\em IEEE signal processing magazine}, 35(4):32--55, 2018.

\bibitem{yang2019deep}
Wenming Yang, Xuechen Zhang, Yapeng Tian, Wei Wang, Jing-Hao Xue, and Qingmin
  Liao.
\newblock Deep learning for single image super-resolution: A brief review.
\newblock {\em IEEE Transactions on Multimedia}, 21(12):3106--3121, 2019.

\end{thebibliography}
\end{document}